\documentclass[aps,prl,twocolumn,superscriptaddress,groupedaddress]{revtex4-1}  
\usepackage{hyperref}
\usepackage{amsfonts}
\usepackage{bbm}
\usepackage{graphicx} 
\usepackage{dcolumn}   
\usepackage{bm}      
\usepackage{amssymb, amsmath}  
\usepackage{enumerate}
\usepackage{slashed}
\usepackage{simplewick}
\usepackage{comment}
\usepackage{color}
\usepackage{soul}

\newcommand{\ba}{\begin{align}}
\newcommand{\ee}{\end{equation}}
\newcommand{\be}{\begin{equation}}

\def\12{\frac{1}{2}}

\newcommand{\en}{\end{align}}

\usepackage{color}

\def\_{{\:\!}}

\def\beq{\begin{equation}}
\def\eeq{\end{equation}}
\def\bea{\begin{eqnarray}}
\def\eea{\end{eqnarray}}

\begin{document}

\setcounter{secnumdepth}{6}

\title{Spectral gaps and mid-gap states in random quantum master equations}
\author{Tankut Can$^1$, Vadim Oganesyan$^{1,2}$, Dror Orgad$^{3}$, Sarang Gopalakrishnan$^{1,2}$}
 \affiliation{$^1$Initiative for the Theoretical Sciences, The Graduate Center, CUNY, New York, NY 10012, USA \\ $^2$Department of Physics and Astronomy, College of Staten Island, Staten Island, NY 10314, USA \\ $^3$Racah Institute of Physics, The Hebrew University, Jerusalem 91904, Israel}

\begin{abstract}

We discuss the decay rates of chaotic quantum systems coupled to noise. We model both the Hamiltonian and the system-noise coupling by random $N \times N$ Hermitian matrices, and study the spectral properties of the resulting Lindblad superoperator.
We consider various random-matrix ensembles, and find that for all of them the asymptotic decay rate remains nonzero in the thermodynamic limit, i.e., the spectrum of the superoperator is gapped as $N \rightarrow \infty$.
A sharp spectral transition takes place as the dissipation strength is increased: for weak dissipation, the non-zero eigenvalues of the master equation form a continuum; whereas for strong dissipation, the asymptotic decay rate is an \emph{isolated eigenvalue}, i.e., a ``mid-gap state'' that is sharply separated from the continuous spectrum of the master equation.
For finite $N$, the probability of finding a very small gap vanishes algebraically with a scaling exponent that is \emph{extensive} in system size, and depends only on the symmetry class of the random matrices and the number of independent decay channels.
We comment on experimental implications of our results.

\end{abstract}

\date{\today}

\maketitle

The approach of a non-equilibrium system to a steady state is a central topic in many-body dynamics. For a system coupled to an external
environment, the rate of approach to the steady state depends nontrivially on the system-environment coupling, i.e., the dissipation strength.
Weak dissipation enhances decay, but strong dissipation can suppress it through the quantum Zeno effect~\cite{weissbook, Zeno, qze_expt}. These
phenomena have been extensively studied, both theoretically and experimentally, for specific models~\cite{syassen, ott2013, diehl2008, cirac2010,
ProsenPRL2008, prosen2008, pz2009, znidaric2010, Kessler2012, Minganti2018, Diehl-Zeno}. Here, in contrast, we discuss them within the
generic setting of random matrix theory (RMT)~\cite{rmtbook}. Historically, RMT was introduced to describe complex dynamical systems
for which a microscopic description would be intractable~\cite{Wigner55,Dyson1962}. RMT is believed to describe the generic long-time
behavior of chaotic quantum systems~\cite{Berry1985,Dymarsky2018a,Schiulaz2018}.
It predicts universal features in the density of states and level statistics that have been verified numerically and experimentally
in many settings~\cite{guhr1998random}. RMT has also been extended to open quantum systems mainly via scattering theory,
where it has captured universal features of the scattering matrix and resonance width distributions of open chaotic systems~\cite{Beenakker97,Schomerus2016}, and has been used to model evolution under effective non-Hermitian Hamiltonians~\cite{nonhermitian}. Given this background, it is somewhat surprising that up to very recently \cite{Poles2018} RMT has not been applied to the Lindblad master equation~\cite{Lindblad1976,GKS,Gorini1978}, which is the standard framework used for exploring open quantum
systems in a wide range of fields from quantum optics to mesoscopics.

\begin{figure}[!t]
\begin{center}
\includegraphics[width =0.45\textwidth,clip=true]{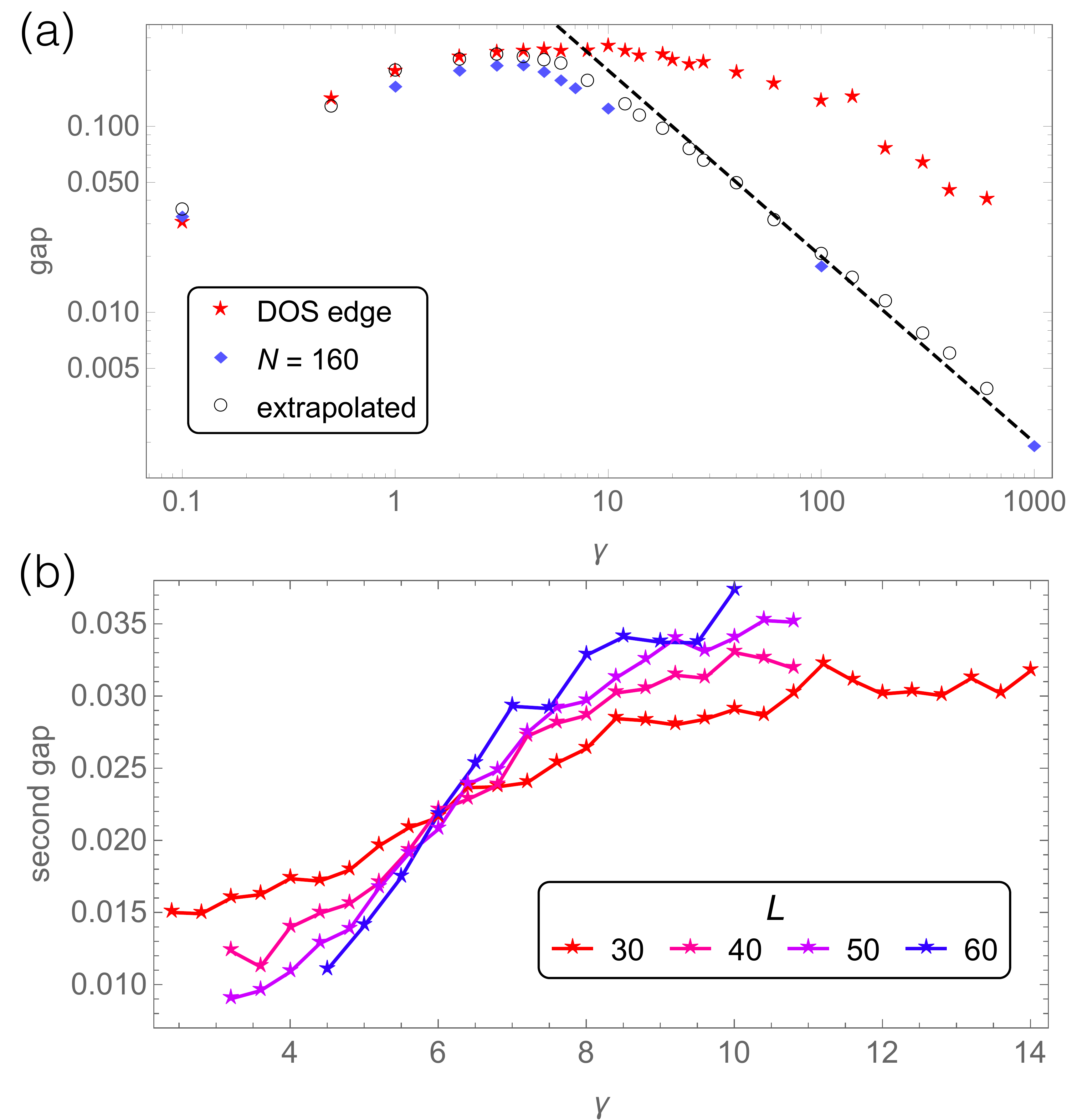}
\caption{(a)~Spectral gap and the edge of the continuum spectrum as a function of dissipation $\gamma$. The gap is computed both
for the largest accessible systems $N = 160$ and by finite-size extrapolation from smaller sizes. The results agree for all $\gamma$
and scale as $(\gamma/2)^{\pm 1}$ for weak/strong dissipation. For $\gamma \geq 6$ the edge of the continuum deviates from the gap,
and the eigenvalues closest to zero become isolated. (b) Finite-size scaling analysis of the spectral transition. The emergence of
a midgap state can be detected by measuring the gap between the first two ``excited'' eigenvalues of the Liouvillian, as discussed
in the main text.}
\label{fig1}
\end{center}
\end{figure}

While much is known about the steady states of specific Lindblad master equations, and although some notions of universality
have been developed for these states and the phase transitions between them~\cite{Kessler2012,shad}, the question of
universality in the \emph{dynamics} of master equations, e.g., their approach to the steady state, has been relatively
unexplored (see however Refs.~\cite{hc,Znidaric2015,Medvedyeva2014,Cai2013}). In this context, it is the aim of the
present work to identify universal dynamical properties of a class of open quantum systems, namely, those that are coupled
to classical white noise, or equivalently to purely dephasing Markovian baths~\cite{Znidaric2013,strunz,joynt}, by
applying RMT to their master equations. Specifically, it addresses the \emph{asymptotic decay rates}
of such purely dephasing master equations.

\begin{figure}[tb]
\begin{center}
\includegraphics[width = 0.45\textwidth]{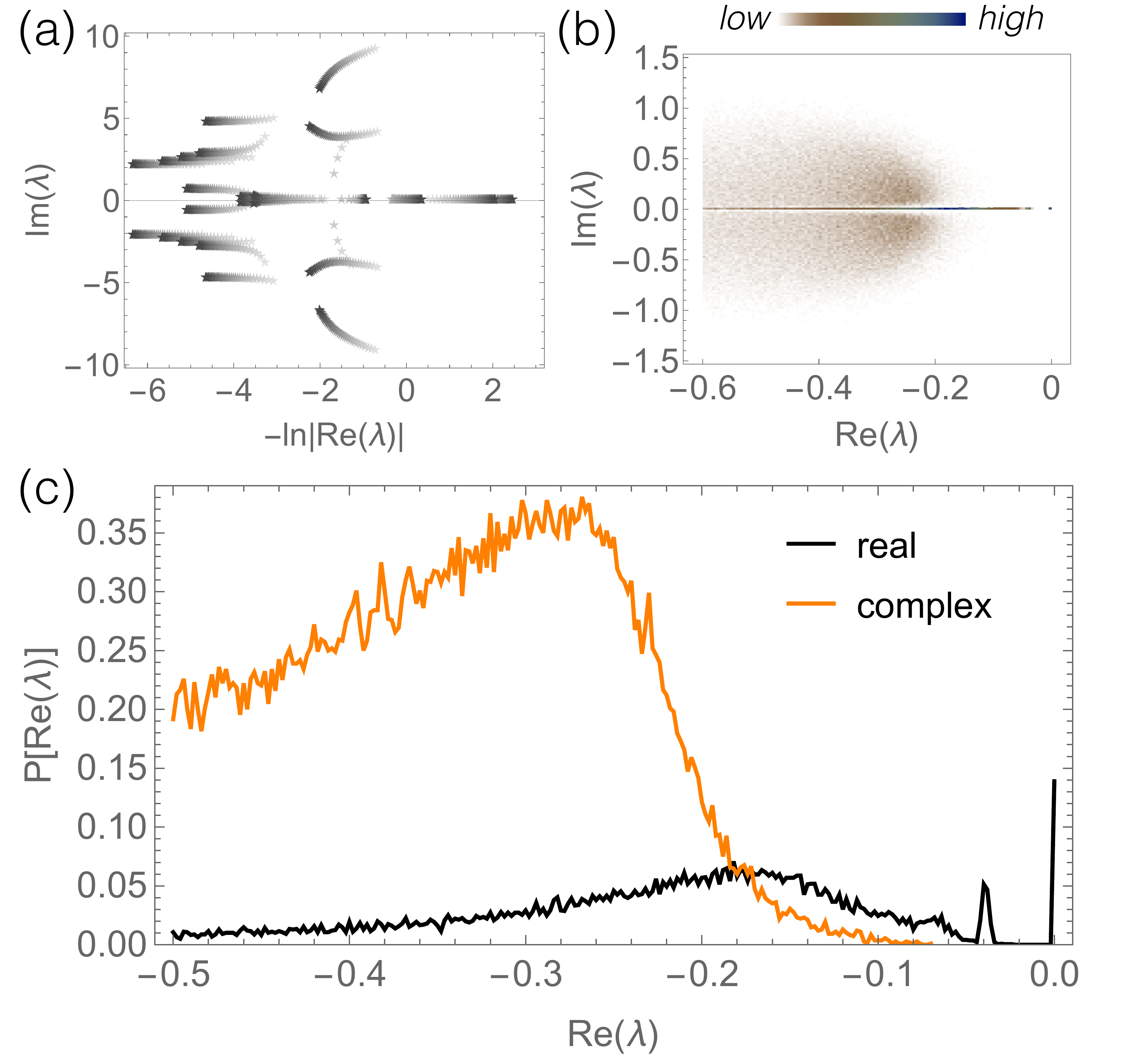}
\caption{(a)~Motion of levels in a fixed realization for $N = 5$ as $\gamma$ is increased from $4$ (lightest) to $20$ (darkest). The $L$-coherences (see main text) decay faster while the $L$-populations decay more slowly. Exceptional points are generically present. (b)~Eigenvalue density in the complex plane for $N = 60, \gamma = 40$, averaged across $100$ realizations, for $|\lambda| < 1$. Note the enhanced density along the real line for small $|\lambda|$. (c)~Histogram of the \emph{real} part of the eigenvalues for the same parameters: purely real eigenvalues are plotted in black and complex ones in orange. The eigenvalues close to threshold are mostly real (derived from $L$-populations), although the two distributions overlap.}
\label{histo}
\end{center}
\end{figure}

\begin{figure}[t]
\begin{center}
\includegraphics[width = 0.4\textwidth]{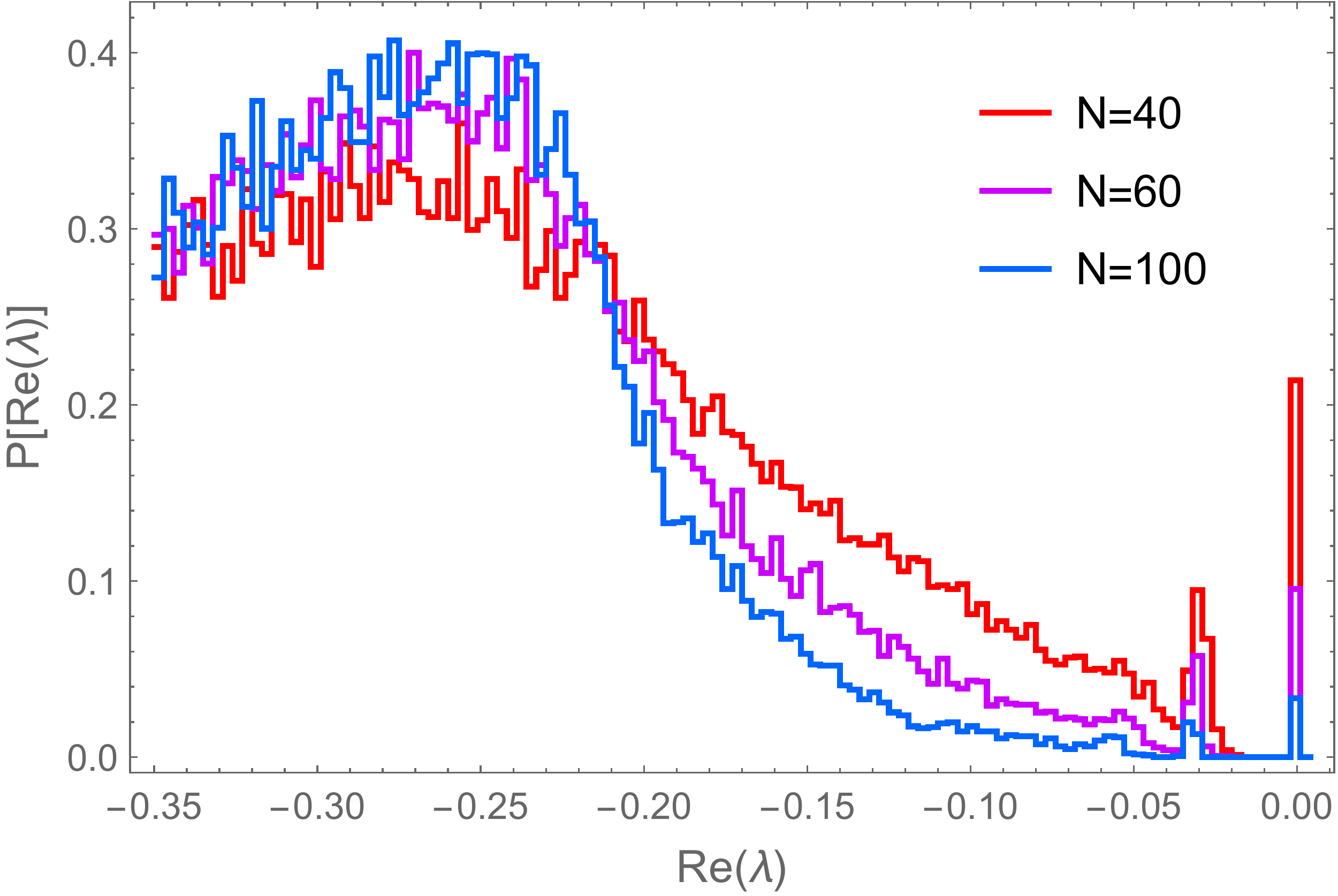}
\caption{Finite-size scaling analysis of the spectral edge for $\gamma = 50$. The presence of a well-defined crossing point allows one to extract the edge of the continuous spectrum. The midgap state appears as a bump in this finite-size data.}
\label{fss}
\end{center}
\end{figure}

The decay rates are the real parts of the eigenvalues of the Liouvillian superoperator $\mathcal{L}$ (defined below), which
governs the evolution of the system's density matrix.  The steady state is a zero mode of $\mathcal{L}$, and the asymptotic
decay rate is determined by the eigenvalue closest in real part to the steady state. For any dissipation, we find that $\mathcal{L}$ is gapped
in the large-$N$ limit, i.e., decay rates sufficiently close to zero become infinitely improbable. However, the structure of the
spectrum near its edge is drastically different at weak and strong dissipation. At weak dissipation, the spectrum consists of an
isolated eigenvalue (the steady state) and a continuum, whereas at strong dissipation isolated ``midgap'' states occur between
the continuum and the steady state (see Figs. \ref{fig1}-\ref{fss}). The asymptotic decay rate (and possibly the next few slowest rates)
are among these midgap states. There is a sharp spectral transition at which the midgap states first appear. The existence of
this transition does not appear to have been noticed in previous work, although related phenomena have been observed in the
mathematical literature on classical dissipative systems~\cite{fpo1, fpo2}.
In addition to establishing these features of the large-$N$ limit, we comment on the universal finite-$N$ scaling of the gap probability distribution, and on global spectral features.

\emph{Master equation}.---We consider the Lindblad equation

\beq\label{mastereq}
\partial_t \rho = \mathcal{L}\rho \equiv -i[H, \rho] + \gamma \sum_{k = 1}^{n_d} \!\left[ L_k \rho L_k^\dagger
- \frac{1}{2} \{ L_k^\dagger L_k, \rho \} \right]\!,
\eeq
where $\rho$ is the system's density matrix, $H$ is its Hamiltonian, $L_k$ are ``jump operators'' representing the coupling of the system to
its environment, and $\gamma$ is the dissipation strength. We focus on the case where the Hamiltonian is an $N \times N$ random matrix from the Gaussian Orthogonal Ensemble (GOE), whose elements have zero mean and variance $\langle H_{ij} H_{kl}\rangle = \frac{1}{2N} (\delta_{ik}\delta_{jl} + \delta_{il}\delta_{jk})$, 
and where there is a single jump operator ($n_d = 1$), statistically independent of $H$ and also drawn from the GOE.
Other cases will be addressed briefly at the end (and in \cite{tc}).
Note that we have scaled the distribution such that the $N\rightarrow\infty$ spectra of $H$ and $L$ reside in $[-\sqrt{2},\sqrt{2}]$.
We concentrate on the large-$N$ limit, but also discuss some nonperturbative features that arise at finite $N$.

\emph{General properties}.---Interpreting the right-hand side of Eq.~\eqref{mastereq} as a ``superoperator'' acting on $\rho$, one can
view the master equation as an eigenvalue problem, $\mathcal{L}\rho = \lambda \rho$. Eq.~\eqref{mastereq} corresponds to a completely
positive trace preserving map, and for $\gamma \ge 0$ all eigenvalues $\lambda$ satisfy ${\rm Re}\, \lambda \leq 0$, with at least one eigenvalue at $\lambda = 0$ (the steady state). For the model considered here, the steady state is unique and given by the infinite temperature thermal state proportional to the identity.

Furthermore, the evolution $\mathcal{L}$ takes legitimate density matrices to other legitimate density matrices.
This requires it to preserve both trace and Hermiticity. It then follows that $(\mathcal{L}\rho)^\dagger = \mathcal{L}(\rho^\dagger)$
and as an $N^2 \times N^2$ matrix its components obey the relation $\mathcal{L}_{ji\_lk} = \mathcal{L}^*_{ij\_kl}$. This ``conjugation''
symmetry can be used to show that the eigenvalues of $\mathcal{L}$ come in complex conjugate pairs~\cite{suppmat}.
Thus, for even (odd) $N$ we expect an even (odd) number of purely real eigenvalues. The number of purely real eigenvalues may vary upon
changing $\gamma$. Complex conjugate pairs of eigenvalues can collide at an ``exceptional point'' and become purely real~\cite{kato_book},
or vice versa [see Fig.~\ref{histo}(a)]. A nontrivial property of the model is that at least $N$ eigenvalues are required to be purely real.
This property follows from combining the conjugation symmetry noted above with the additional symmetry of the Lindblad superoperator $\mathcal{L}_{ij\_kl} = \mathcal{L}_{kl\_ij}$ which follows when $H$ and $L$ are real symmetric matrices~\cite{suppmat}.

\emph{Weak dissipation}.---In the absence of dissipation the dynamics is governed by the operator $-i [H, \cdot]$.
The Liouvillian eigenvalues correspond to differences, $i(E_\alpha - E_\beta)$,
between Hamiltonian eigenenergies. This spectrum has $N$ eigenstates of the form $|\alpha\rangle \langle \alpha|$
with eigenvalue zero, and $N(N-1)$ states of the form $|\alpha \rangle \langle \beta|$ with imaginary nonzero eigenvalues that come
in complex conjugate pairs. In the language of nuclear magnetic resonance, the former correspond to ``populations'' and the
latter to ``coherences''~\cite{nmrbook}. Level repulsion in the spectrum of $H$ manifests itself as a suppression of
the density of eigenvalues with small nonzero imaginary parts $|\mathrm{Im}(\lambda)| \alt 1/N$. 

When $\gamma$ is small, we can treat it perturbatively, and to first order find that the populations and coherences are decoupled.

For the populations, one obtains a symmetric classical master equation acting on the $N \times N$ subspace. The off-diagonal terms of this master equation are $\gamma|L_{\alpha\beta}|^2$ with $L_{\alpha\beta} = \langle \alpha|L |\beta\rangle$, while the diagonal terms are determined by the constraint that each column should sum to zero. Note that all the terms in this matrix are sign-definite, so one can disorder-average the master equation to get a matrix element (i.e., a transition rate) $\langle |L_{\alpha\beta}|^2 \rangle \sim 1/(2N)$ between any pair of populations.
Consequently, the spectrum of this averaged classical master equation contains a unique steady state and $N-1$ degenerate states with eigenvalue $\lambda = -\gamma/2$. Meanwhile, at first order and up to $1/N$ corrections, each coherence $|\alpha\rangle\langle\beta|$ picks up a (real)
perturbative shift of the form $-(\gamma/2) \sum\nolimits_\eta (|L_{\alpha\eta}|^2 + |L_{\beta\eta}|^2) = -\gamma/2$. Therefore, to linear order in $\gamma$,  $N^2 - 1$ states rigidly move away from the steady state by an amount $-\gamma/2$.
Since there is a hard gap at first order in $\gamma$, we expect it to be robust, provided that higher-order perturbative corrections do not diverge in the large-$N$ limit. We have checked this to order $\gamma^2$~\cite{suppmat}.

Our results imply that for small $\gamma$, the dynamics is captured by the mean-field master equation
\begin{align}
\partial_{t}\rho = - i [ H, \rho] - \frac{\gamma }{2  } \left[  \rho - \frac{1}{N} \, \mathbbm{1} + \frac{1}{N} ( \rho - \rho^{*}) \right].\label{mean-field-liouv}
\end{align}
To leading order in $1/N$, the mean-field dissipator is the generator of a depolarizing channel \cite{Kastoryano2013}. 
This approximation fails, however, to yield the $O(\gamma^{4})$ variance of the decay rates at large $N$, as reveled 
by perturbation theory~\cite{suppmat}.

\emph{Strong dissipation}.---We now turn to the case of strong dissipation. To leading order in $1/\gamma$
the eigenvalues of $\mathcal{L}$ are squared differences in the spectrum of the jump operator $L$, i.e.,
if $\kappa_a$ are the $L$-eigenvalues then $\lambda=-(\gamma/2) (\kappa_a - \kappa_b)^2$. Here too, there are $N$ ``diagonal'' zero modes of the form $|a\rangle \langle a|$;
we call these ``$L$-populations''. We call the remaining eigenstates, of the form $|a \rangle \langle b|$, ``$L$-coherences''.
In this limit the spectrum is entirely real and gapless, with bandwidth $4\gamma$ and level spacing $\sim \gamma/N^2$.

We now perturb in the Hamiltonian, taking $N \rightarrow \infty$ at large but finite $\gamma$. At first order,
an $L$-coherence $|a\rangle \langle b|$ picks up a purely imaginary $O(1/\sqrt{N})$ shift
$\tau_{ab} = i (\langle b | H | b \rangle - \langle a | H | a \rangle)$, with the result
\beq
\lambda_{ab} = -\frac{\gamma}{2} (\kappa_a - \kappa_b)^2 + i \tau_{ab}.
\eeq
The $L$-populations, however, remain exact zero modes to first order.
To resolve these degeneracies, one must go to second order in the Hamiltonian where populations and
coherences mix, but one can disentangle the subspaces via a Schrieffer-Wolff transformation~\cite{Kessler}.
Each $L$-population connects to $2(N-1)$ coherences, obtained by changing one (but not both) of its indices.
Consequently, any two populations, $|a\rangle \langle a|$ and $|b\rangle \langle b|$ are connected by the
second order matrix element
\beq\label{matrel}
\varphi_{ab} = - \frac{|H_{ab}|^2}{\lambda_{ab}} - \frac{|H_{ab}|^2}{\lambda^*_{ab}} =\frac{4}{\gamma} \frac{|H_{ab}|^2 (\kappa_a - \kappa_b)^2}{(\kappa_a - \kappa_b)^4 + (2/\gamma)^{2}\tau_{ab}^2}.
\eeq

One may worry that since many [$O(N^{3/4})$] of the $L$-coherences that are coupled to a population
have eigenvalues that are smaller than a typical $H_{ab}$, the coherences are strongly coupled to the populations
and there is no way to separate them. However, note that the matrix elements~\eqref{matrel} vanish
for pairs of $L$-populations with very similar eigenvalues $\kappa_a \approx \kappa_b$. Consequently,
coherences with small $|\lambda_{ab}|$ do not couple strongly to the populations and the required
separation between the two sets can be carried out to second order in $H$~\cite{suppmat}.

In the subspace of $L$-populations the spectrum consists of two types of states. The lowest few eigenstates are isolated
and remain separate from the continuum even after disorder averaging. Their form and their eigenvalues $\lambda_n=-(2/\gamma)n$,
with integer $n\geq 0$, can be obtained analytically~\cite{suppmat}.
For more negative $\lambda$, the peaks corresponding to individual eigenvalues become more closely spaced as well as broader,
and merge into a continuum upon disorder averaging~\cite{suppmat}. At the same time, in the subspace of $L$-coherences,
non-degenerate second-order perturbation theory shifts states near $\lambda = 0$ by an amount $-c/\gamma$~\cite{suppmat},
with $c$ a constant, resulting in an overall gapped spectrum. Numerically, we find $c>2$ such that the $\lambda_1$ state 
(with possibly additional $n>1$ levels) always appears to the right of the continuum, leading to the large-$\gamma$ behavior $\Delta=2/\gamma$, 
in accord with the quantum Zeno effect.

\emph{Numerical results}.---We now present numerical results on the distribution of gaps at various values of $\gamma$ and $N$.
Since the results obtained by finite-size extrapolation agree well with those obtained for the largest manageable systems
($N = 160$) using the Arnoldi method, we expect that the calculated gap values are close to the thermodynamic limit.
Both methods give a transition from $\Delta=\gamma/2$ at weak dissipation to $\Delta=2/\gamma$ at large dissipation with
a maximum around $\gamma\approx 3.5$, see Fig. \ref{fig1}a.

We used finite-size scaling to locate the edge of the continuous spectrum of $\mathcal{L}$. Within our numerical constraints we were
able to detect, in the range $0.1 \leq \gamma \leq 100$, sharpening of the density of states (DOS) edge with system size,
giving a well-defined crossing point (Fig.~\ref{fss}). This is consistent with the conjecture that the DOS jumps discontinuously
at the edge of the spectrum. When $\gamma \alt 6$ the jump in the DOS occurs at the gap, but for larger $\gamma$ the jump is at
a larger value of $|\lambda|$ than the gap. In this regime, it seems that there is at least one isolated state,
and possibly multiple states, at decay rates appreciably lower than the edge of the continuous spectrum.
These can clearly be seen as isolated peaks in the DOS for the larger system sizes in Fig.~\ref{fss}.

To better locate this transition, we computed the distance between the second- and third-highest eigenvalues (i.e., the second and third lowest in absolute value) of $\mathcal{L}$ as a function of $\gamma$. A clear flow reversal is seen [Fig.~\ref{fig1}(b)], indicating a phase transition. For $\gamma \alt 6$ these two eigenvalues approach each other with increasing system size, as one expects for continuum states. For larger $\gamma$, however, these two eigenvalues move \emph{away} from each other with increasing system size, as the DOS jump steepens with $N$ and the isolated eigenvalue becomes more clearly separated from the continuum. This flow reversal allows us to identify the spectral transition point as $\gamma_c \approx 6$.

\begin{figure}[t]
\begin{center}
\includegraphics[width = 0.45\textwidth]{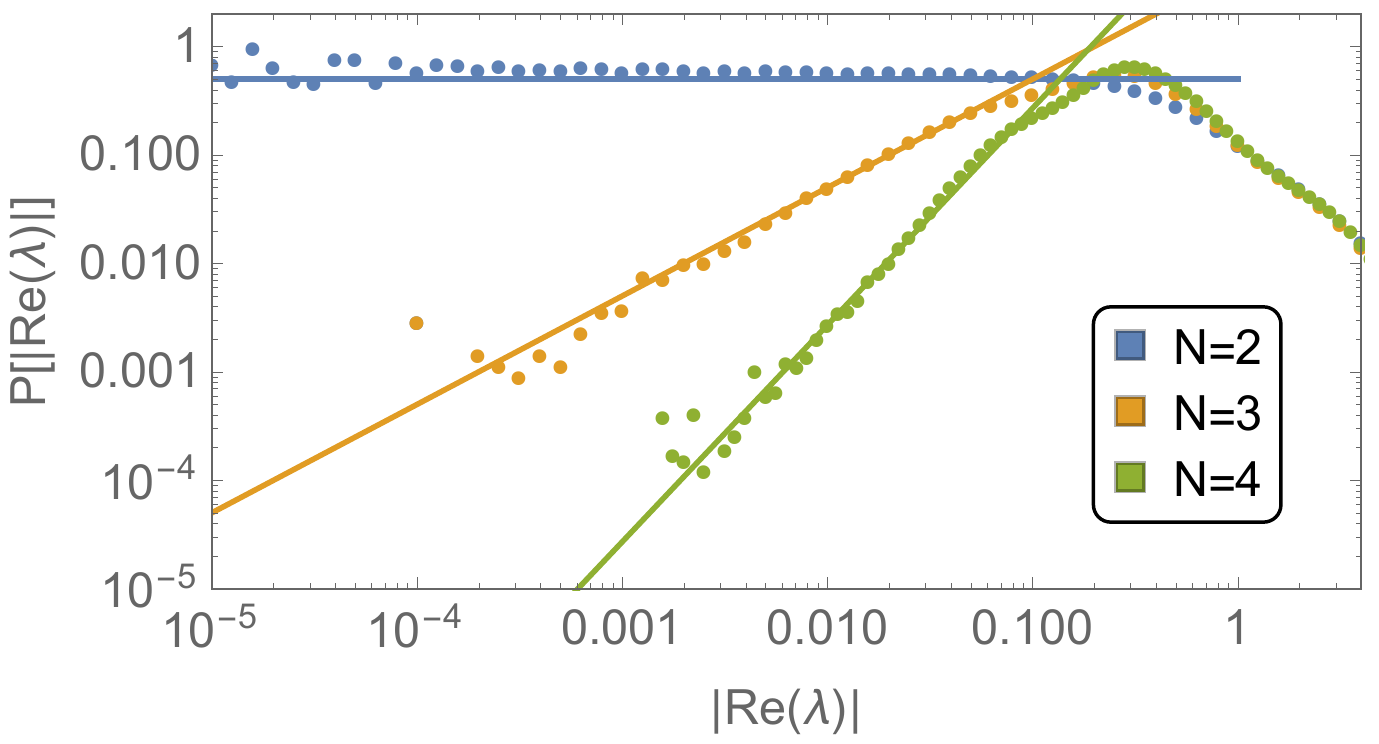}
\caption{Eigenvalue density close to $\lambda = 0$ at $\gamma = 2$ for small system sizes, $N = 3,5,7$, averaged over $500,000$ realizations; the behavior is consistent with Eq.~\eqref{scaling} (straight lines).}
\label{smallN}
\end{center}
\end{figure}

\emph{Finite-size corrections}.---Our results indicate that as $N \rightarrow \infty$ the disorder-averaged density of states is zero at sufficiently small $|\lambda|$. We now discuss how this hard gap sets in, by considering finite-size systems. We argue on general grounds that the density of states as $\lambda \rightarrow 0$ is sensitive to the symmetry classes of $H$ and $L_k$, as well as to the size of the Hilbert space $N$ and the number of distinct jump operators $L_k$; however, its behavior is independent of $\gamma$. For the case of $k$ jump operators, the probability of finding a gap $\Delta$ scales as

\beq\label{scaling}
P(\Delta) \sim \Delta^{\frac{\beta k}{2} (N - 1) - 1},
\eeq
where $\beta = 1,2,4$ for orthogonal, unitary, and symplectic ensembles. This prediction is compared with numerics for the orthogonal ensemble
(Fig.~\ref{smallN}), where results for other ensembles are shown in~\cite{suppmat}. The figures plot the asymptotic density of eigenvalues,
and show that it compares favorably with the scaling behavior, Eq. (\ref{scaling}). When the Hamiltonian and jump operators belong to
different ensembles, the appropriate value of $\beta$ is that corresponding to the lower-symmetry (higher $\beta$) ensemble.

Eq.~\eqref{scaling} can be understood via a golden rule argument. For simplicity we first treat the case of a real (orthogonal-class)
Hamiltonian and  a single real Hermitian jump operator $L$. Consider a system initialized in the eigenstate $|n\rangle$ of the Hamiltonian.
In the presence of a Markovian bath (which has an energy-independent density of states) its decay rate is given by
$\Gamma_n = \sum\nolimits_{m\ne n} |\langle n | L | m \rangle|^2$,
where $m$ runs over all other eigenstates of $H$. There are $N - 1$ matrix elements in this expression, and (given that $H$ and $L$ are mutually uncorrelated and taken from the GOE) $L_{nm} \equiv \langle n | L | m \rangle$ can be regarded as independent real Gaussian random numbers. For $\Gamma_n$ to be small we need all matrix elements to be small. One can approximate the cumulative probability distribution
\beq
P(\Gamma_n \leq \Delta) \sim \prod\nolimits_m P(L_{nm}^2 \leq \Delta) \propto (\sqrt{\Delta})^{N - 1},
\eeq
from which Eq.~\eqref{scaling} follows by differentiation.
The other cases are simple generalizations. For example, in the unitary ensemble, each matrix element is a complex number,
and is only small when its real and imaginary parts are separately small, yielding a factor of two in the exponent.
In general, therefore, the disorder-averaged DOS at a given small $\lambda$ vanishes exponentially in $N$.

\emph{Discussion}.---In this work we have explored the spectral structure of the low-lying (i.e., long-lived) states of a generic dephasing Lindblad master equation. Such master equations describe chaotic quantum dots, as well as, e.g., systems of trapped ions with long-range interactions subject to electromagnetic noise. The Hamiltonian can generically be modeled as a random matrix. Whether the system-bath coupling (which sets the jump operators $L$) should have random-matrix or Poisson level statistics is not clear in general. However, none of our results relied on level repulsion in either $H$ or $L$, and we expect that they will continue to apply so long as the eigenvectors of $L$ look like random vectors in the eigenbasis of $H$ and vice versa. We found that the master equation is gapped in the large-$N$ limit. However, a spectral transition occurs as a function of $\gamma$: for $\gamma \agt 6$, the asymptotic decay rate is set by a mid-gap state whose eignevalue is well below
the edge of the continuum (Fig.~\ref{fig1}).

This unexpected feature has direct implications for relaxation from a generic initial state. This relaxation takes place in two stages: there is a shorter timescale on which the decay is relatively complex, and is set by the continuum density of states, as well as an intermediate-time regime with a slower exponential decay set by the eigenvalue of the mid-gap state.

This work has focused on the case of the orthogonal ensemble because of its simplicity as well as its direct relation to classical noisy dynamics. A companion paper \cite{tc} finds similar features in the Ginibre ensemble (i.e., for non-Hermitian dissipators with no symmetries), for which a more systematic calculation is possible. A natural question is how far these results extend to other random matrix ensembles, as well as to systems with spatial structure, such as banded random matrices. We will address this question in a subsequent work.

\bigskip
{\it Note Added:} While this manuscript was under preparation, the preprint \cite{Poles2018} appeared which discusses random Lindblad generators. This work considers the spectrum of the dissipator with $N^{2} -1$ jump operators with no symmetries, and no Hamiltonian. They similarly find a spectral gap, and are able to characterize the full eigenvalue spectral density on the complex plane.
\bigskip

\begin{acknowledgments}
S.~G. acknowledges support from NSF Grant No. DMR-1653271. V.~O. and D.~O. acknowledge support by
the United States-Israel Binational Science Foundation (Grant No. 2014265). V.~O. acknowledges 
support from NSF Grant No. DMR-1508538.
\end{acknowledgments}

\bibliography{randlind}

\end{document}


\title{Supplemental Material for: "Spectral gaps and mid-gap states in random quantum master equations"}

\author{Tankut Can$^1$, Vadim Oganesyan$^{1,2}$, Dror Orgad$^{3}$, Sarang Gopalakrishnan$^{1,2}$}
 \affiliation{$^1$Initiative for the Theoretical Sciences, The Graduate Center, CUNY, New York, NY 10012, USA \\ $^2$Department of Physics and Astronomy, College of Staten Island, Staten Island, NY 10314, USA \\ $^3$Racah Institute of Physics, The Hebrew University, Jerusalem 91904, Israel}

\maketitle

In this supplementary document, we provide a detailed discussion of the symmetries of the master equation and of its eigenvalue distribution
in the perturbative small and large $\gamma$ limits. We also present supplemental numerical results on the eigenvalue distribution and on finite-size gaps.

\section{Symmetries and perturbation theory}

\subsection{Statement of the problem}

We consider the quantum master equation for the density matrix $\rho$ in the case of a single Hermitian jump operator
\begin{equation}
\label{eq:quantummaster}
\partial_t\rho=-i[H,\rho]-\gamma\left[\frac{1}{2}\{L^2,\rho\}-L\rho L \right]=- i [H, \rho] - \frac{\gamma}{2}[L,[L,\rho]].
\end{equation}
Furthermore, we assume that the Hamiltonian $H$ and the jump operator $L$ are represented by $N\times N$
real symmetric random matrices drawn from the GOE ensemble, \ie,
\begin{equation}
\label{eq:GOE}
P(H)\propto \exp\left(-\frac{N}{2}\Tr\left[H^2\right]\right), \quad \langle H_{ij} H_{kl} \rangle = \frac{1}{2N} (\delta_{ik}\delta_{jl} + \delta_{ik}\delta_{jl}),
\end{equation}
and similarly for $L$. Note that we have scaled the variance of the probability distribution (\ref{eq:GOE})
by the dimension of the Hilbert space, $N$, such that in the large-$N$ limit the spectrum of $H$ and $L$ resides
within the segment $[-\sqrt{2},\sqrt{2}]$.

The master equation (\ref{eq:quantummaster}) can be written in terms of a Lindbladian superoperator, $\L$,
represented by an $N^2\times N^2$ matrix with composite indices $ij$ and $kl$, acting on $\rho$
\begin{equation}
\label{eq:masterL}
\partial_t\rho_{ij}=\sum_{kl}\L_{ij\_kl}\,\rho_{kl},
\end{equation}
with
\begin{equation}
\label{eq:Lcomp}
\L_{ij\_kl}=-[iH+\frac{\gamma}{2}L^2]_{ik}\delta_{jl}
+\delta_{ik}[iH-\frac{\gamma}{2}L^2]_{jl}+\gamma L_{ik}L_{jl}.
\end{equation}

For Hermitian jump operators, the fact that the dissipator can be written as a nested commutator implies that the Lindblad superoperator 
has the structure
\begin{align}
\mathcal{L} = - i C_{H} - \frac{\gamma}{2} C_{L}^{2}\;,	\label{eq:lind-rep}
\end{align}
where $C_{M} = M \otimes \mathbbm{1} - \mathbbm{1} \otimes M$ is the superoperator representation of the commutator with $M$. When $M$ is Hermitian, $C_{M}$ is also Hermitian.

\medskip

Our goal is to explore the spectral properties of $\L$ as function of the dissipation strength $\gamma$.
In particular, we are interested in its spectral gap, as defined below, which governs the slowest decay of
the system towards a steady state.

\subsection{Properties and representation of $\L$}
\label{subsec:prop}

The Lindblad superoperator has in general the following properties which we make use of:
\begin{enumerate}[leftmargin=0.4cm]
\item{}$\L_{ij\_kl}=\L^*_{\widebar{ij}\_\widebar{kl}}$, where $\widebar{ij}=ji$. Thus, $\L_{ij\_\widebar{ij}}$ is real. In the language of linear maps, this is equivalent to the condition that the Lindbladian preserves Hermiticity $\left( \mathcal{L}[\rho]\right)^{\dagger} = \mathcal{L}[ \rho^{\dagger}]$.

\item{} The eigenvalues of $\L$ are either real or come in complex conjugated pairs. This follows from the item above, since if $\rho$ is an eigenmode such that $\mathcal{L}[\rho] = \lambda \rho$, then $\rho^{\dagger}$ is also an eigenmode satisfying $\mathcal{L}[\rho^{\dagger}] = \lambda^{*} \rho^{\dagger}$. In the superoperator representation, this property is a consequence of the symmetry $[\L,{\cal C}{\cal K}]=0$, where in the block representation introduced in Eq. (\ref{eq:Lparam}) below
\begin{equation}
\label{eq:Cdef}
{\cal C}=\left[\begin{array}{ccc}
1&0&0\\
0&0&1\\
0&1&0
\end{array}\right],
\end{equation}
and ${\cal K}$ is the complex conjugation operator. As a result, if $\rho$ is an eigenvector of $\L$
with eigenvalue $\lambda$, then ${\cal C}{\cal K}\rho$ is an eigenvector with eigenvalue $\lambda^*$.

\item{} The eigenvalues of $\L$ have a non-positive real part. While it is generally true, this property is easiest to show when $L$ is a Hermitian matrix, which is the case we consider in this paper. Let $\mathcal{L} | \rho \rangle  = \lambda | \rho \rangle $, where $|\rho\rangle$ is the vectorized eigenmode of the Liouvillian. Then utilizing the representation (\ref{eq:lind-rep}), the real part of the eigenvalue satisfies
\begin{align}
{\rm Re} \lambda  = - \frac{\gamma}{2} \frac{ \langle \rho |C_{L}^{2} | \rho \rangle}{ \langle \rho | \rho \rangle} \le 0	, \quad {\rm for} \, \gamma \ge 0.
\end{align}
The inequality follows because $C_{L}^{2}$, as the square of a Hermitian matrix, is clearly positive-semidefinite. We also see that convergence requires $\gamma \ge 0$.

\item{} The Lindblad equation is trace preserving, which means that ${\rm tr} \mathcal{L}[\rho] = 0$. Using the tensor representation (\ref{eq:masterL}), this implies  $\sum_{i} \mathcal{L}_{ii kl} = 0$ for all $k,l$.

\end{enumerate}
\bigskip
\bigskip
With the additional assumption that $H$ and $L$ are {\bf real symmetric matrices}, we have the following:
\bigskip
\begin{enumerate}[leftmargin=0.4cm]
\setcounter{enumi}{4}
\item{}The Lindblad superoperator becomes symmetric: $\L_{ij\_kl}=\L_{kl\_ij}$.

\item{} The steady state $\mathcal{L}[\rho_{ss}] = 0$ is the infinite temperature thermal state $\rho_{ss} = \frac{1}{N} \mathbbm{1}$. More generally, this is true when $L$ are normal matrices.  Eq. (\ref{eq:Lcomp}) implies that
if $[H,L]=0$ such that they share a basis of $N$ simultaneous eigenvectors $v^\alpha$, then
$\rho_{ij}^\alpha=v_i^\alpha v_j^\alpha$ constitute $N$ zero modes.

\item{} We find it useful to order the composite indices of $\L$ in the following way
\begin{equation}
\label{eq:Lparam}
\begin{centering}
\begin{array}{c}
\hspace{53pt}
\!\!\!\!\!\!\!\!\!\!\!\!\!
\begin{array}{c c c}
\rule{0pt}{21pt}
\makebox[25pt][c]{$\vcenter{$\,kk$\vspace{10pt}}$}&
\makebox[50pt][c]{$\vcenter{$\,kl$\vspace{10pt}}$}&
\makebox[50pt][c]{$\vcenter{$\,\widebar{kl}$\vspace{10pt}}$}
\end{array}\\
\!\!\!\!\!\!\!\!\!\!\!{\cal L}=
\begin{array}{c}
\rule{0pt}{21pt}
\makebox[25pt][c]{$\vcenter{$ii$\vspace{11pt}}$}\\
\rule{0pt}{42pt}
\makebox[25pt][c]{$\vcenter{$\ij$\vspace{21pt}}$}\\
\rule{0pt}{42pt}
\makebox[25pt][c]{$\vcenter{$\widebar{ij}$\vspace{20pt}}$}
\end{array}
\!\!\left[
\begin{array}{c|c|c}
\rule{0pt}{21pt}
\makebox[25pt][c]{$\vcenter{$A$\vspace{10pt}}$}&
\makebox[50pt][c]{$\vcenter{$B$\vspace{10pt}}$}&
\makebox[50pt][c]{$\vcenter{$\;B^*$\vspace{10pt}}$}\\
\hline
\rule{0pt}{42pt}
\makebox[25pt][c]{$\vcenter{$B^T$\vspace{22pt}}$}&
\makebox[50pt][c]{$\vcenter{$C$\vspace{20pt}}$}&
\makebox[50pt][c]{$\vcenter{$D$\vspace{20pt}}$} \\
\hline
\rule{0pt}{42pt}
\makebox[25pt][c]{$\vcenter{$B^\dagger$\vspace{22pt}}$}&
\makebox[50pt][c]{$\vcenter{$\;D^*$\vspace{20pt}}$}&
\makebox[50pt][c]{$\vcenter{$\;C^*$\vspace{20pt}}$}
\end{array}
\right]
\end{array},
\vspace{5pt}
\end{centering}
\end{equation}
where $i<j$ and $k<l$. As a consequence of properties 1 and 5 one finds that:
\newline
$A$ is a real symmetric $N\times N$ matrix, containing the "populations",
\newline
$B$ is a complex $N\times N(N-1)/2$ matrix, and
\newline
$C$ is a complex symmetric $N(N-1)/2\times N(N-1)/2$ matrix, which together with
the complex Hermitian $N(N-1)/2\times N(N-1)/2$ matrix $D$, contains the "coherences".

In this representation the steady state is   $\rho_{ss}=(\overbrace{1/N,\cdots,1/N}^N,\overbrace{0,\cdots,0}^{N(N-1)})^T$.

\item{}If all eigenvalues of $\L$ are distinct, as is typically expected based on the randomness of $H$
and $L$ (and in the absence of any additional symmetries), then it is diagonalizable, \ie, $\L=V\Lambda V^{-1}$. Here, $V$ is a matrix whose columns are the
eigenvectors of $\L$, and $\Lambda$ is a diagonal matrix containing the corresponding eigenvalues.
Since $\mathcal{L}$ is symmetric, the eigenvectors $\rho^\alpha$ can be made an orthonormal
basis with respect to the inner product $\sum_{ij}\rho^\alpha_{ij}\rho^\beta_{ij}=\delta^{\alpha\beta}$,
and that for this choice $V^{-1}=V^T$. Note
that $\L$ may still be diagonalizable even in the presence of degeneracy as demonstrated by the case $[H,L]=0$.

\item{} $\L$ is guaranteed to have at least $N$ real eigenvalues. This fact is a consequence of a
theorem by Carlson\cite{Carlson65}, stating that a necessary and sufficient condition for a complex
matrix $M$ to have at least $m$ real eigenvalues is the existence of a Hermitian matrix ${\cal C}$
with $|\sigma({\cal C})|=m$, such that $M{\cal C}$ is also Hermitian. Here, $\sigma$ denotes the signature.
In our case, ${\cal C}$ is given by Eq. (\ref{eq:Cdef}) and $\sigma({\cal C})=N$.

\end{enumerate}

\subsection{The small $\gamma$ limit}

In the limit of weak dissipation the dynamics is largely governed by the Hamiltonian, while $L$ acts as a
small perturbation. For this reason we choose to analyze $\L$ in the eigenbasis of $H$,
where $H_{ij}=\epsilon_i\delta_{ij}$ and the components of $\L$ take the form
 \begin{eqnarray}
\label{eq:Asmall}
&&A_{ii\_jj}=-\gamma\left[{L^2}_{ij}\delta_{ij}-(L_{ij})^2\right], \\
\label{eq:Bsmall}
&&B_{ii\_kl}=-\frac{\gamma}{2}\left[{L^2}_{ik}\delta_{il}+\delta_{ik}{L^2}_{il}-2L_{ik}L_{il}\right], \\
\label{eq:Csmall}
&&C_{ij\_kl}=i(\epsilon_j-\epsilon_i)\delta_{ik}\delta_{jl}-\frac{\gamma}{2}\left[{L^2}_{ik}\delta_{jl}
+\delta_{ik}{L^2}_{jl}-2L_{ik}L_{jl}\right], \\
\label{eq:Dsmall}
&&D_{ij\_\widebar{kl}}=-\frac{\gamma}{2}\left[{L^2}_{il}\delta_{jk}+\delta_{il}{L^2}_{jk}-2L_{il}L_{jk}\right].
\end{eqnarray}

We will first analyze the spectrum of $A$ and of the matrix $F=\left[\begin{array}{cc} C & D\\ D^* & C^*\end{array}\right]$
separately, and then will consider the effects of their coupling through $B$.

\subsubsection{The spectrum of $A$}

Within the GOE ensemble, Eq. (\ref{eq:GOE}), the elements of $H$ and $L$ are normally-distributed independent random variables
with zero mean $\mu(L_{ij})=0$, and standard deviation $\sigma(L_{ij})=1/\sqrt{(2-\delta_{ij})N}$. The central limit theorem
then implies that in the large-$N$ limit the diagonal elements of $A_{ii}=-\gamma\sum_{j\neq i}(L_{ij})^2$ are normally distributed,
with slight dependence between them ($L_{ij}$ appears both in $A_{ii}$ and $A_{jj}$) and
\begin{equation}
\label{eq:disAii}
\mu(A_{ii})=-\gamma\frac{N-1}{2N}\simeq -\frac{\gamma}{2}, \;\;\;\;\;\;\;\;
\sigma(A_{ii})=\gamma\sqrt{\frac{N-1}{2N^2}}\simeq\frac{\gamma}{\sqrt{2N}}.
\end{equation}
The off-diagonal elements are chi-squared distributed with
\begin{equation}
\label{eq:disAij}
\mu(A_{ij})=\frac{\gamma}{2N},  \;\;\;\;\;\;\;\;
\sigma(A_{ij})=\frac{\gamma}{\sqrt{2}N},
\end{equation}
and are dependent on the diagonal elements $A_{ii}$.

\smallskip

Next, we decompose $A$ according to
\begin{equation}
\label{eq:Adecomp}
A=W+\frac{\gamma}{2N}K-\frac{\gamma}{2}I,
\end{equation}
where $K$ is the constant matrix $K_{ij}=1$. The elements of $W$ are distributed in the same way
as the elements of $A$, except that their mean is shifted to zero $\mu(W_{ij})=0$. Because
$\sum_i W_{ij}=\sum_j W_{ij}=0$ one finds $[W,K]=0$, and the eigenvectors of $A$ are simultaneous
eigenvectors of $W$ and $K$ (and trivially of $I$). Among them the zero mode $v^0=(1/N,\cdots,1/N)$
is always present and the remaining eigenvectors $v^\alpha$, $\alpha=1,\cdots, N-1$, are orthogonal
to it and thus satisfy $\sum_i v^\alpha_i=0$. Consequently, $Kv^\alpha=0$ and their eigenvalues are
$-\gamma/2+w^\alpha$, where $w^\alpha$ are the eigenvalues with respect to $W$. Since the elements of
$W$, as those of $A$, are not independent the eigenvalue distribution deviates from
Wigner's semicircle law. Nevertheless, we can estimate its width. To this end, consider the eigenvalue
problem $\sum_j W_{ij}v^\alpha_j=w^\alpha v^\alpha_i$, which implies $\mu(w^\alpha)=0$, provided any
correlations between $v^\alpha_i$ and $W_{ij}$ are neglected. Squaring the eigenvalue equation,
summing over $i$ and using the normalization of wavevectors leads to
\begin{equation}
\label{eq:Weigs}
(w^\alpha)^2=\sum_{ij}W_{ij}^2(v^\alpha_j)^2+\sum_i\sum_{j\neq k}W_{ij}W_{ik}v^\alpha_j v^\alpha_k.
\end{equation}
Neglecting the dependence of $v^\alpha_i$ on $W_{ij}$ one obtains
\begin{equation}
\label{eq:Awidth}
\mu\left[(w^\alpha)^2\right]=\left[\mu\left(W_{jj}^2\right)+\sum_{i\neq j}\mu\left(W_{ij}^2\right)\right]
\sum_j \left(v_j^\alpha\right)^2=\frac{\gamma^2}{2N}+(N-1)\frac{\gamma^2}{2N^2}\simeq\frac{\gamma^2}{N}.
\end{equation}
Hence, we conclude that the spectrum of $A$ is narrowly distributed around $-\gamma/2$ with a width that
scales as $\gamma/\sqrt{N}$. This conclusion is supported by our numerics.
\smallskip

The matrix $A$ is also known as a Markov generator, and was studied in some detail in Refs. \onlinecite{Timm, Bryc}. In particular, it was proven in Ref. \onlinecite{Bryc} using the method of moments that the limiting eigenvalue distribution as $N \to \infty$ is given by the sum of two delta functions: one at the origin with unit weight corresponding to the steady state, and one at $ - \gamma/2$ with weight $N-1$.

\subsubsection{The spectrum of $F$}

Since $\gamma$ is small our strategy is to estimate the eigenvalue distribution of $F$ by treating
its off-diagonal elements as a perturbation. The diagonal elements are
\begin{equation}
\label{eq:Pdiagel}
-\frac{\gamma}{2}\left[{L^2}_{ii}+{L^2}_{jj}-2L_{ii}L_{jj}\right]+i(\epsilon_j-\epsilon_i)\equiv x_{ij}+iy_{ij},
\;\;\;\;\;\;\;\;\;\;\;\;\;\;\; (i\neq j).
\end{equation}
Using the fact that in the large-$N$ limit ${L^2}_{ii}$ are normally distributed with $\mu({L^2}_{ii})=1/2$ and
$\sigma({L^2}_{ii})=1/\sqrt{2N}$ and neglecting the $O(1/N)$ contribution from the $L_{ii}L_{jj}$ piece,
we find that the $x_{ij}$s are normally distributed with
\begin{equation}
\label{eq:xdist}
\mu(x)=-\frac{\gamma}{2},  \;\;\;\;\;\;\;\;
\sigma(x)=\frac{\gamma}{2\sqrt{N}}.
\end{equation}
On scales larger than the mean level spacing $\delta\sim 1/N$ one can largely neglect the correlations
between the $\epsilon_i$s and evaluate the distribution of $y_{ij}$ using Wigner's semicircle law
\begin{equation}
\label{eq:ydist}
P(y_{ij})
=\frac{1}{\pi^2}\int_{-\sqrt{2}}^{\sqrt{2}} d\epsilon\sqrt{2-\epsilon^2}\sqrt{2-(y_{ij}-\epsilon)^2}
\,\Theta\left(2\sqrt{2}-|y_{ij}|\right)=f(y_{ij})\,\Theta\left(2\sqrt{2}-|y_{ij}|\right),
\end{equation}
with
\begin{equation}
\label{eq:fdef}
f(y)=\frac{2\sqrt{2}+|y|}{6\pi^2}\left\{
(8+y^2){\cal E}\left[\left(\frac{2\sqrt{2}-|y|}{2\sqrt{2}+|y|}\right)^2\right]
-2^{5/2}|y|{\cal K}\left[\left(\frac{2\sqrt{2}-|y|}{2\sqrt{2}+|y|}\right)^2\right]\right\},
\end{equation}
where ${\cal E}$ and ${\cal K}$ are the complete elliptic integrals and $\Theta$ is the step function.
The above result fails for $|y_{ij}|\lesssim \delta$ due to level repulsion, and in this regime
$P(|y_{ij}|\lesssim \delta)\sim |y_{ij}|$. However, for the rough estimates that will follow we only need
to note that Eq. (\ref{eq:ydist}) implies that in the large-$N$ limit
\begin{equation}
\label{eq:ymoments}
\mu(y)=0,  \;\;\;\;\;\;\;\;
\sigma(y)=1.
\end{equation}

Let us consider the shift of an unperturbed eigenvalue $x_{ij}+iy_{ij}$ within second order perturbation
theory. The corresponding eigenstate $\rho_{ij}$ is connected by $(N-1)^2-N+2$ off-diagonal elements of $F$
to other states $\rho_{kl}$ with both $k\neq i$ and $l\neq j$ (type $A$), and by $2(N-2)$ off-diagonal elements to
states $\rho_{kl}$ with either $k=i$ or $l=j$ (type $B$). We begin by examining the type $A$ elements,
which are of the form $z_{ij\_kl}=\gamma L_{ik}L_{jl}$, with the distribution function
\begin{equation}
\label{eq:P(z)}
P(z)=\frac{N}{\pi}\int_{-\infty}^\infty dl \frac{1}{\gamma|x|}e^{-Nl^2}e^{-N z^2/\gamma^2 l^2}
=\frac{2N}{\pi\gamma}K_0\left(\frac{2N}{\gamma}|z|\right),
\end{equation}
where $K_0$ is the modified Bessel function. Consequently, the numerator of the corresponding term in
the eigenvalue shift
\begin{equation}
\label{eq:pshift}
\frac{z_{ij\_ kl}^2}{x_{ij}-x_{kl}+i(y_{ij}-y_{kl})}\equiv\frac{\zeta_{ij\_ kl}}{x_{ij\_ kl}+iy_{ij\_ kl}},
\end{equation}
is distributed according to
\begin{equation}
\label{eq:Pzeta}
P^A_{\zeta}(\zeta)=\frac{2N}{\pi\gamma}\frac{1}{\sqrt{\zeta}}K_0\left(\frac{2N}{\gamma}\sqrt{\zeta}\right)\Theta(\zeta).
\end{equation}
We are particularly interested in the shift of the real part of the eigenvalue due to these terms
\begin{equation}
\label{eq:realshiftA}
\Delta x^A_{ij}=\sum_{k\neq i, l\neq j}\frac{x_{ij\_ kl}}{x_{ij\_ kl}^2+y_{ij\_ kl}^2}\zeta_{ij\_ kl}
\equiv\sum_{k\neq i, l\neq j} \xi_{ij\_ kl}\zeta_{ij\_ kl}\equiv\sum_{k\neq i, l\neq j} \chi_{ij\_ kl},
\end{equation}
as the unperturbed real part is very narrowly distributed in the large-$N$ limit, see Eq. (\ref{eq:xdist}).
To this end we note that $x_{ij\_ kl}=x_{ij}-x_{kl}$ is normally distributed with $\mu(x)=0$
and $\sigma(x)=\gamma/\sqrt{2N}$. To simplify the analysis we neglect the weak correlations between
$y_{ij}$ and $y_{kl}$ and approximate the distribution of $y_{ij\_ kl}=y_{ij}-y_{kl}$ by a normal distribution
with $\mu(y)=0$ and $\sigma(y)=\sqrt{2}$, in accordance with Eq. (\ref{eq:ymoments}). We have checked numerically
that this is a fair approximation. Under these assumptions
\begin{eqnarray}
\label{eq:P(xi)}
\nonumber
P^A_{\xi}(\xi)&=&\frac{N^{1/2}}{2\pi\gamma}\int_{-\infty}^\infty dx dy\, e^{-Nx^2/\gamma^2} e^{-y^2/4}
\delta\left(\xi-\frac{x}{x^2+y^2}\right)\\
&=&\frac{N^{1/2}}{2\pi\gamma}\frac{1}{|\xi|^3}\int_{-\infty}^\infty dy \frac{1}{(1+y^2)^2}
e^{-N/[\gamma^2\xi^2(1+y^2)^2]} e^{-y^2/[4\xi^2(1+y^2)^2]}.
\end{eqnarray}
By considering the behavior of the integrand in different regimes it is possible to approximate
\begin{equation}
\label{eq:P(xi)approx}
P^A_{\xi}(\xi)\sim \left\{
\begin{array}{cc}
\frac{N^{1/2}}{\gamma}|\xi|^{-3} & \;\;\;\;\;\;|\xi|>\frac{N^{1/2}}{\gamma} \\
\frac{\gamma^{1/2}}{N^{1/4}}|\xi|^{-3/2} & \;\;\;\;\;\;\frac{N^{1/2}}{\gamma}>|\xi|>\frac{\gamma}{N^{1/2}} \\
\frac{N^{1/2}}{\gamma} & \;\;\;\;\;\;\frac{\gamma}{N^{1/2}}>|\xi|
\end{array}
\right. .
\end{equation}
Combining Eqs. (\ref{eq:Pzeta},\ref{eq:P(xi)approx}) allows us to estimate the distribution of
$\chi_{ij\_ kl}=\xi_{ij\_ kl}\zeta_{ij\_ kl}$
\begin{equation}
\label{eq:P(chi)}
P^A_{\chi}(\chi)=\int_0^\infty d\zeta\, \frac{1}{\zeta} P^A_{\zeta}(\zeta) P^A_{\xi}\left(\frac{\chi}{\zeta}\right)
\sim \left\{
\begin{array}{cc}
\frac{\gamma^3}{N^{7/2}}|\chi|^{-3} & \;\;\;\;\;\;|\chi|>\frac{\gamma}{N^{3/2}} \\
\frac{\gamma^{3/2}}{N^{5/4}}|\chi|^{-3/2} &
\;\;\;\;\;\;\frac{\gamma}{N^{3/2}}>|\chi|>\frac{\gamma^3}{N^{5/2}} \\
\frac{N^{5/2}}{\gamma^3}+\frac{N^{5/4}}{\gamma^{3/2}}|\chi|^{-1/2}\ln\left(\frac{\gamma^3}{N^{5/2}|\chi|}\right) & \
\;\;\;\;\;\;\frac{\gamma^3}{N^{5/2}}>|\chi|
\end{array}
\right. ,
\end{equation}
which implies $\mu(|\chi^A|)\equiv\mu(|\chi_{ij\_kl}|)\sim \gamma^2/N^2$.

\smallskip

Consider now the type $B$ terms with $i=k$ or $j=l$. The latter, for example,
are dominated by $z_{ij\_kj}=-(\gamma/2){L^2}_{ik}$, which is normally distributed with $\mu(z)=0$ and
$\sigma(z)=\gamma/(4\sqrt{N})$. Consequently, the numerators of the corresponding perturbative
correction
\begin{equation}
\label{eq:realshiftB}
\Delta x^B_{ij}=\sum_{k\neq i}\frac{x_{ij\_ kj}}{x_{ij\_ kj}^2+y_{ij\_ kj}^2}\zeta_{ij\_ kj}+
\sum_{l\neq j}\frac{x_{ij\_ il}}{x_{ij\_ il}^2+y_{ij\_ il}^2}\zeta_{ij\_ il}
\equiv\sum_{k\neq i} \xi_{ij\_ kj}\zeta_{ij\_ kj}+\sum_{l\neq j} \xi_{ij\_ il}\zeta_{ij\_ il}
\equiv\sum_{k\neq i} \chi_{ij\_ kj}+\sum_{l\neq j} \chi_{ij\_ il},
\end{equation}
are distributed according to
\begin{equation}
\label{eq:Pzetadi}
P^B_{\zeta}(\zeta)=\frac{4\sqrt{N}}{\sqrt{2\pi}\gamma}\frac{1}{\sqrt{\zeta}}e^{-(8N/\gamma^2)\zeta}\,\Theta(\zeta).
\end{equation}
The real part of the denominator $x_{ij\_kj}\approx -(\gamma/2)({L^2}_{ii}-{L^2}_{kk})$ is normally distributed with
$\mu(x)=0$ and $\sigma(x)=\gamma/(2\sqrt{N})$. We approximate the distribution of the imaginary part
$y_{ij\_kj}=\epsilon_k-\epsilon_i$, Eq. (\ref{eq:ydist}), by a normal distribution with $\mu(y)=0$
and $\sigma(y)=1$. Consequently, the distribution $P^B_\xi(\xi)$ of $\xi_{ij\_kj}$ and $\xi_{ij\_il}$ shares the same
approximate begavior of $P^A_\xi(\xi)$, as given by Eq. (\ref{eq:P(xi)approx}), and for $\chi_{ij\_kj}$, $\chi_{ij\_jl}$
we may estimate
\begin{equation}
\label{eq:P(chi)di}
P^B_{\chi}(\chi)=\int_0^\infty d\zeta\, \frac{1}{\zeta} P^B_{\zeta}(\zeta) P^B_{\xi}\left(\frac{\chi}{\zeta}\right)
\sim \left\{
\begin{array}{cc}
\frac{\gamma^3}{N^{3/2}}|\chi|^{-3} & \;\;\;\;\;\;|\chi|>\frac{\gamma}{N^{1/2}} \\
\frac{\gamma^{3/2}}{N^{3/4}}|\chi|^{-3/2} &
\;\;\;\;\;\;\frac{\gamma}{N^{1/2}}>|\chi|>\frac{\gamma^3}{N^{3/2}} \\
\frac{N^{3/4}}{\gamma^{3/2}}|\chi|^{-1/2} &
\;\;\;\;\;\;\frac{\gamma^3}{N^{3/2}}>|\chi|
\end{array}
\right. ,
\end{equation}
implying $\mu(|\chi^B|)\equiv\mu(|\chi_{ij\_kj}|)=\mu(|\chi_{ij\_il}|)\sim \gamma^2/N$.

\smallskip

Finally, ignoring the dependence of $|\chi_{ij\_ kl}|$ on the
position of the unperturbed levels within the spectrum we may approximate
\begin{eqnarray}
\label{eq:rshiftapprx}
\nonumber
\mu(\Delta x_{ij})&=&\mu(\Delta^A x_{ij})+\mu(\Delta^B x_{ij}) \\
&\sim &\mu(|\chi^A|)\sum_{k\neq i,l\neq j}{\rm sign}(\chi_{ij\_ kl})
+\mu(|\chi^B|)\Bigg[\sum_{k\neq i}{\rm sign}(\chi_{ij\_ kj})+\sum_{l\neq j}{\rm sign}(\chi_{ij\_ il})\Bigg]
\sim \gamma^2{\rm erf}\left(\frac{x_{ij}+\gamma/2}{\gamma/\sqrt{2N}}\right).
\end{eqnarray}
Here we used the fact that owing to the normal distribution of the $x_{ij}$s, a state with a given $x_{ij}$
has a proportion of $1/2\pm (1/2){\rm erf}[(x_{ij}+\gamma/2)/({\gamma/\sqrt{2N}})]$ of the terms in the sum appear
with sign $\pm 1$.

\medskip

\begin{figure}[!!!h]
\begin{center}
\includegraphics[width = 0.532\textwidth]{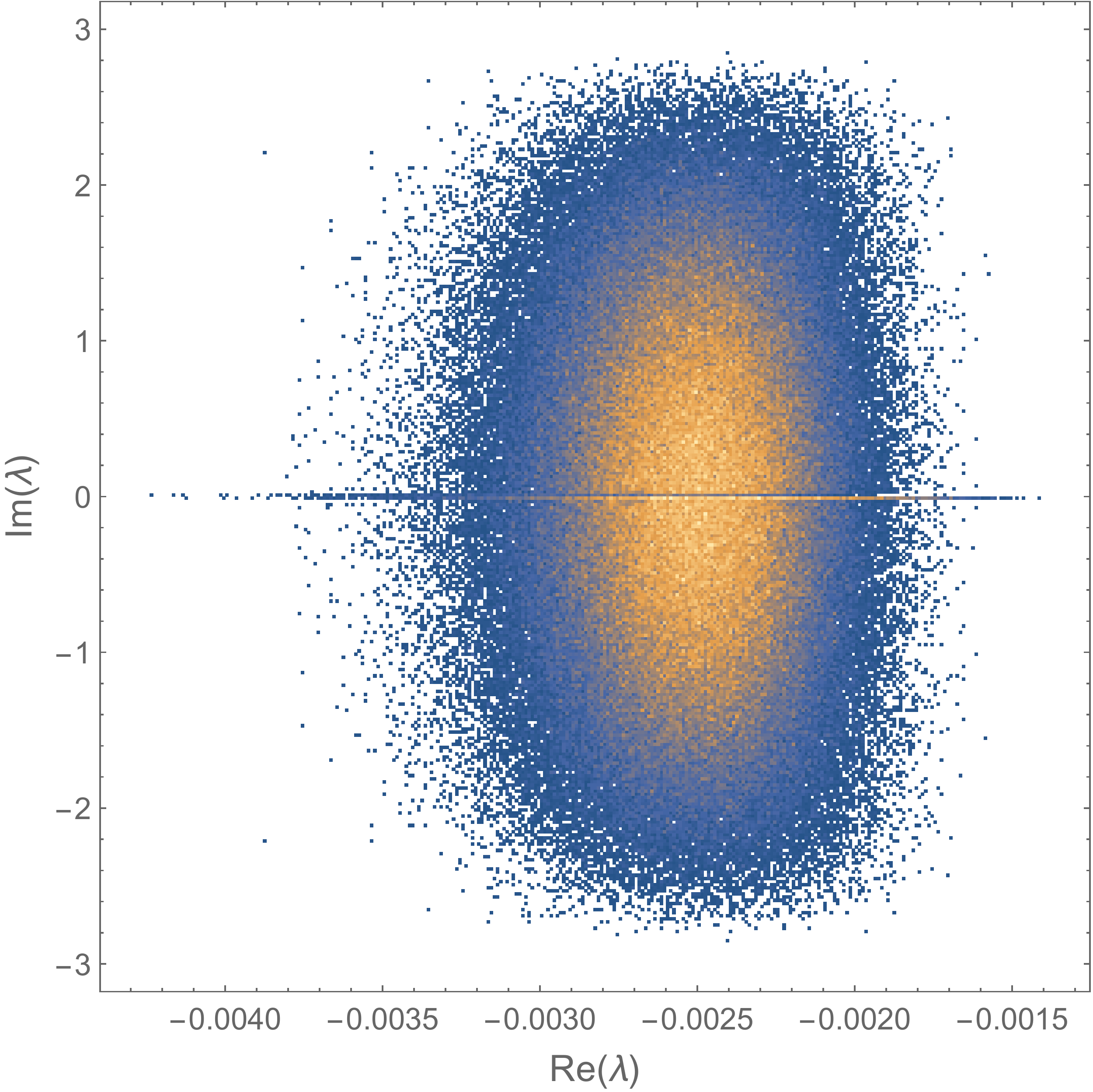}
\includegraphics[width = 0.45\textwidth]{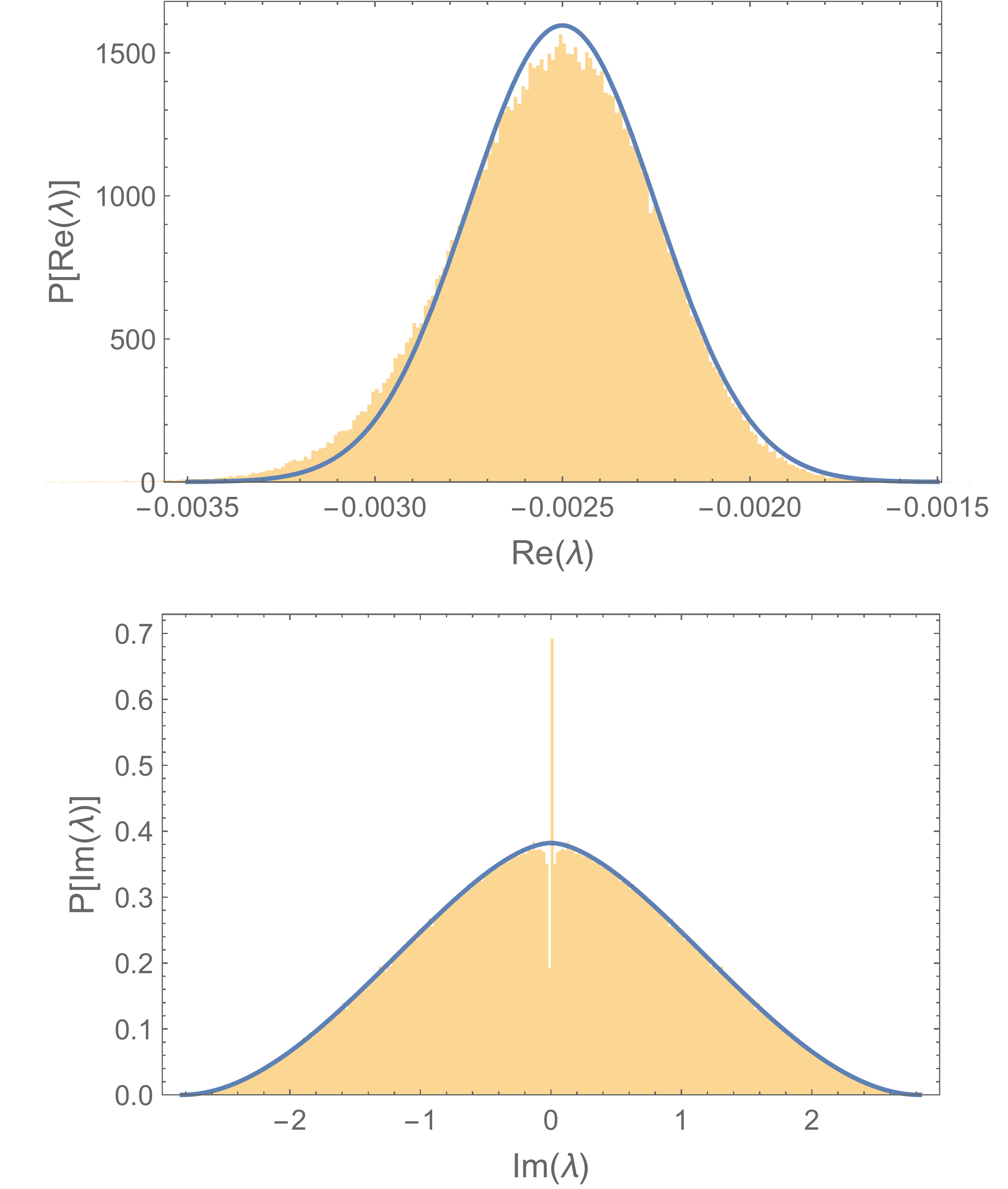}
\end{center}
\vspace{-0.5cm} \caption[]{\small Left: Eigenvalue distribution of $\L$, binned from 40 realizations
for the case $\gamma=0.005$ and $N=100$. Not shown is the zero-mode at the origin. Right: Projections of the
distribution on the real and imaginary axes. The former is concentrated around $-\gamma/2$ and for the range of parameters
used its standard deviation is $\gamma/(2\sqrt{N})$. The blue line depicts a normal distribution with these parameters.
We expect that as $N\rightarrow\infty$ the standard deviation transitions to $O(\gamma^2)$.
The imaginary parts are distributed with standard deviation 1, and the blue line depict the distribution given
by Eq. (\ref{eq:ydist}). Note the peak due to the real eigenvalues inside the dip around zero that is inherited from the
level repulsion of $H$.}
\label{fig-smallgamma}
\end{figure}

As a rough estimate for the distribution of the real part of the eigenvalues $\tilde{x}_{ij}=x_{ij}+\Delta x_{ij}$
we replace $\Delta x_{ij}$ by its mean, Eq. (\ref{eq:rshiftapprx}), and approximate ${\rm erf}(x)\sim x$ to
obtain $\tilde{x}_{ij}=x_{ij}+c\gamma\sqrt{N}(x_{ij}+\gamma/2)$, where $c$ is a constant of order 1. The resulting
$\tilde{x}_{ij}$ is normally distributed with $\mu(\tilde{x})=-\gamma/2$ and a standard deviation that evolves from
$\gamma/(2\sqrt{N})$ for $N\ll 1/\gamma^2$ to $\sigma(\tilde{x})=c\gamma^2/2$ for $N\gg 1/\gamma^2$, see Fig. \ref{fig-smallgamma}.
We note that Eqs. (\ref{eq:P(chi)}) and (\ref{eq:P(chi)di}) imply that $\sigma(|\chi^{A,B}|)$ diverges
logarithmically due to the behavior of $P_\chi$ at large values. This would lead to positive values of $\tilde{x}_{ij}$
in contradiction to their non-positiveness (see property 3 of section \ref{subsec:prop}), indicating the failure of
second order perturbation theory for the edges of the distribution. The same remark also holds for the normal
distribution estimated above.

\bigskip
Consider now the shift in the imaginary part of the eigenvalue due to the $A$-type terms
\begin{equation}
\label{eq:imshiftA}
\Delta y^A_{ij}=-\sum_{k\neq i, l\neq j}\frac{y_{ij\_ kl}}{x_{ij\_ kl}^2+y_{ij\_ kl}^2}\zeta_{ij\_ kl}
\equiv\sum_{k\neq i, l\neq j} \eta_{ij\_ kl}\zeta_{ij\_ kl}\equiv\sum_{k\neq i, l\neq j} \nu_{ij\_ kl},
\end{equation}
Under the same assumptions used before
\begin{eqnarray}
\label{eq:P(eta)}
\nonumber
P^A_{\eta}(\eta)&=&\frac{N^{1/2}}{2\pi\gamma}\int_{-\infty}^\infty dx dy\, e^{-Nx^2/\gamma^2} e^{-y^2/4}
\delta\left(\eta+\frac{y}{x^2+y^2}\right)\\
&=&\frac{N^{1/2}}{2\pi\gamma}\frac{1}{|\eta|^3}\int_{-\infty}^\infty dx \frac{1}{(1+x^2)^2}
e^{-Nx^2/[\gamma^2\eta^2(1+x^2)^2]} e^{-1/[4\eta^2(1+x^2)^2]}.
\end{eqnarray}
By considering the behavior of the integrand in different regimes it is possible to approximate
\begin{equation}
\label{eq:P(eta)approx}
P^A_{\eta}(\eta)\sim \left\{
\begin{array}{cc}
\frac{N^{1/2}}{\gamma}|\eta|^{-3} & \;\;\;\;\;\;|\eta|>\frac{N^{1/2}}{\gamma} \\
|\eta|^{-2} & \;\;\;\;\;\;\frac{N^{1/2}}{\gamma}>|\eta|>1 \\
\frac{\gamma^2}{N} & \;\;\;\;\;\;1>|\eta|
\end{array}
\right. .
\end{equation}
Combining Eqs. (\ref{eq:Pzeta},\ref{eq:P(eta)approx}) we can estimate
\begin{equation}
\label{eq:P(nu)}
P^A_{\nu}(\nu)=\int_0^\infty d\zeta\, \frac{1}{\zeta} P^A_{\zeta}(\zeta) P^A_{\eta}\left(\frac{\nu}{\zeta}\right)
\sim \left\{
\begin{array}{cc}
\frac{\gamma^3}{N^{7/2}}|\nu|^{-3} & \;\;\;\;\;\;|\nu|>\frac{\gamma}{N^{3/2}} \\
\frac{\gamma^{2}}{N^{2}}|\nu|^{-2} &
\;\;\;\;\;\;\frac{\gamma}{N^{3/2}}>|\nu|>\frac{\gamma^2}{N^2} \\
\frac{N}{\gamma}|\nu|^{-1/2}\left[1+\ln\left(\frac{\gamma^2}{N^2|\nu|}\right)\right] & \
\;\;\;\;\;\;\frac{\gamma^2}{N^2}>|\nu|
\end{array}
\right. .
\end{equation}
Numerically, its seems that the decay in the intermediate region $\gamma/N^{3/2}>|\nu|>\gamma^2/N^2$ is slightly slower
than $1/\nu^2$. This would eliminate the logarithmic correction to $\mu(|\nu^A|)\equiv\mu(|\nu_{ij\_kl}|)$ and lead to
$\mu(|\nu^A|)\sim \gamma^2/N^2$.

\smallskip

The shift due to the $B$ type terms is
\begin{equation}
\label{eq:imaglshiftB}
\Delta y^B_{ij}=-\sum_{k\neq i}\frac{y_{ij\_ kj}}{x_{ij\_ kj}^2+y_{ij\_ kj}^2}\zeta_{ij\_ kj}-
\sum_{l\neq j}\frac{y_{ij\_ il}}{x_{ij\_ il}^2+y_{ij\_ il}^2}\zeta_{ij\_ il}
\equiv\sum_{k\neq i} \eta_{ij\_ kj}\zeta_{ij\_ kj}+\sum_{l\neq j} \eta_{ij\_ il}\zeta_{ij\_ il}
\equiv\sum_{k\neq i} \nu_{ij\_ kj}+\sum_{l\neq j} \nu_{ij\_ il}.
\end{equation}
Due to similar reasons to the ones outlined above, the distribution $P^B_\eta(\eta)$ of $\eta_{ij\_kj}$ and $\eta_{ij\_il}$
shares the same approximate begavior of $P^A_\eta(\eta)$, as given by Eq. (\ref{eq:P(eta)approx}), and for $\nu_{ij\_kj}$,
$\nu_{ij\_jl}$ we may estimate
\begin{equation}
\label{eq:P(nu)di}
P^B_{\nu}(\nu)=\int_0^\infty d\zeta\, \frac{1}{\zeta} P^B_{\zeta}(\zeta) P^B_{\eta}\left(\frac{\nu}{\zeta}\right)
\sim \left\{
\begin{array}{cc}
\frac{\gamma^3}{N^{3/2}}|\nu|^{-3} & \;\;\;\;\;\;|\nu|>\frac{\gamma}{N^{1/2}} \\
\frac{\gamma^2}{N}|\nu|^{-2} &
\;\;\;\;\;\;\frac{\gamma}{N^{1/2}}>|\nu|>\frac{\gamma^2}{N} \\
\frac{N^{1/2}}{\gamma}|\nu|^{-1/2} &
\;\;\;\;\;\;\frac{\gamma^2}{N}>|\nu|
\end{array}
\right. ,
\end{equation}
implying (assuming that the decay in the intermediate region is slightly slower than $\nu^{-2}$, as appears numerically)
that $\mu(|\nu^B|)\equiv\mu(|\nu_{ij\_kj}|)=\mu(|\nu_{ij\_il}|)\sim \gamma^2/N$.

\smallskip

Using these results we may approximate the shift in the imaginary part of the eigenvalues
\begin{equation}
\label{eq:ishiftapprx}
\mu(\Delta y_{ij})=\mu(\Delta^A y_{ij})+\mu(\Delta^y y_{ij})
\sim \mu(|\nu^A|)\sum_{k\neq i,l\neq j}{\rm sign}(\nu_{ij\_ kl})
+\mu(|\nu^B|)\Bigg[\sum_{k\neq i}{\rm sign}(\nu_{ij\_ kj})+\sum_{l\neq j}{\rm sign}(\nu_{ij\_ il})\Bigg].
\end{equation}
Away from the origin the resulting $O(\gamma^2)$ shift is negligible compared to the $O(1)$ width of $P(y_{ij})$.
There is, however, the question of the effect on the behaviour for $|y_{ij}|<1/N$, where the level repulsion of
$H$ implies $P(y_{ij})\sim N|y_{ij}|$. Owing to this behavior, a given state with $|y_{ij}|<1/N$ has a proportion
of $1/2\pm (N/2)y_{ij}^2$ of the terms in the sum appear with sign $\mp y_{ij}/|y_{ij}|$. As a result,
the shift is of order $-N\gamma^2 y_{ij}^2{\rm sign}(y_{ij})$, which can be neglected compared to $y_{ij}$,
and level repulsion persists.

\subsubsection{The effect of the $B$ couplings}

The analysis of the preceding section can be readily applied to the coupling between the $A$ and $F$ sectors via
the $B$ matrix. While the zero mode of $A$ is unaffected, the remaining $N-1$ eigenvalues of $A$ are shifted along
the real axis. Their resulting distribution is approximately normal with $\mu(x)=-\gamma/2$ and $\sigma(x)$ that
varies from $\gamma/\sqrt{N}$ to $O(\gamma^2)$ as $N$ is increased beyond $1/\gamma^2$. Note that they stay real,
since the perturbative corrections come in complex conjugate pairs. At the same time, the coupling through $B$ has
negligible effect on the $F$ spectrum. This is a result of the fact that there are only $N-2$ type-$A$ perturbative
corrections for a given $kl$, and only two type-$B$ corrections (in the original $A$ basis). This leads to a total
correction that scales as $1/N$.

\vspace{0.5cm}

\subsubsection{Laplace Transform and long time limit}

An alternative method to track the gap structure of the eigenvalues within perturbation theory utilizes the Laplace transform of the resolvent of the Lindblad superoperator, which we define directly
\begin{align}
F(t) = \left\langle {\rm tr} \, e^{ \mathcal{L} t} \right\rangle 	 = \left\langle \sum_{a} e^{ \lambda_{a} t} \right\rangle.
\end{align}
Evaluating the eigenvalues to first order in perturbation theory, we can split the sum into two pieces. One controlled by the eigenvalues of $A$, and the others given by the first-order shift of the complex eigenvalues Eq. (\ref{eq:Pdiagel})
\begin{align}
F(t) = \left\langle {\rm tr} e^{ A t} \right\rangle + \left\langle \sum_{i\ne j} e^{ (x_{ij} + i y_{ij}) t}	\right\rangle .
\end{align}
Using the results of Ref. \onlinecite{Bryc}, the first term tends to
\begin{align}
\left\langle {\rm tr} e^{ A t} \right\rangle \to   	1 + (N-1) e^{ - \frac{\gamma}{2} t}.
\end{align}
The second term can be evaluated exactly to yield
\begin{align}
	\left\langle \sum_{i\ne j} e^{ (x_{ij} + i y_{ij}) t}	\right\rangle = (N^{2} - N)\left(K(t) - N\right) \left(1 + \frac{2 \gamma t}{N}\right)^{-1/2} \left( 1 + \frac{\gamma t}{N}\right)^{-1/2} \left( 1 + \frac{\gamma t}{2 N} \right)^{ -  (N-2)} ,	
\end{align}
where $K(t)$ is the spectral form factor of the Hamiltonian $H$. In the limit $N \to \infty$, this becomes exponentially decaying 
$\exp ( - \gamma t / 2)$, with a life-time identical to that produced by $A$.

\subsection{The large $\gamma$ limit}

In the limit of strong dissipation the dynamics is largely governed by the jump operator and we
use its eigenbasis, where $L_{ij}=\kappa_i\delta_{ij}$, to express the components of $\L$ as
 \begin{eqnarray}
\label{eq:Alarge}
&&A_{ii\_jj}=0, \\
\label{eq:Blarge}
&&B_{ii\_ kl}=i\left[\delta_{ik}H_{il}-H_{ik}\delta_{il}\right], \\
\label{eq:Clarge}
&&C_{ij\_kl}=-\frac{\gamma}{2}(\kappa_i-\kappa_j)^2\delta_{ik}\delta_{jl}+
i\left[\delta_{ik}H_{jl}-H_{ik}\delta_{jl}\right], \\
\label{eq:Dlarge}
&&D_{ij\_\widebar{kl}}=i\left[\delta_{il}H_{jk}-H_{il}\delta_{jk}\right].
\end{eqnarray}
Here, we would like to bring $\L$ into a block diagonal form
\begin{equation}
\label{eq:blockdiag}
\L=\left[\begin{array}{cc} A' & 0\\ 0 & F'\end{array}\right],
\end{equation}
where $A'$ in an $N\times N$ Hermitian matrix whose eigenvalues are the $N$ guaranteed real
eigenvalues of $\L$, and $F'$ is a complex symmetric $N(N-1)\times N(N-1)$ matrix. To achieve
this we employ a generalized Schrieffer-Wolff transformation \cite{Kessler}, which to lowest order
on $1/\gamma$ gives
\begin{eqnarray}
\label{eq:swres}
A'&=&-BC_0^{-1}B^T+{\rm c.c.}, \\
F'&=&\left[\begin{array}{cc} C & D\\ D^* & C^*\end{array}\right],
\end{eqnarray}
where $[C_0]_{ij\_kl}=-(\gamma/2)(\kappa_i-\kappa_j)^2\delta_{ik}\delta_{jl}$.

\subsubsection{The spectrum of $A'$}

We are interested in finding the eigenvalues $\Lambda$ and eigenvectors ${\bm v}$ of $A'$.
Using Eq. (\ref{eq:swres}) the eigenvalue equation becomes
\begin{equation}
\label{eq:eigsA1}
\frac{4}{\gamma}\sum_{\substack{j=1\\ j\neq i}}^N(H_{ij})^2\frac{v_j-v_i}{(\kappa_j-\kappa_i)^2}=\lambda v_i.
\end{equation}
In the $N\rightarrow\infty$ the $\kappa$s become dense, with density
\begin{equation}
\label{eq:semicircle}
\nu(\kappa)=\frac{N}{\pi}\sqrt{2-\kappa^2}\,\Theta(2-\kappa^2),
\end{equation}
and it is useful to parameterize the eigenvector components not by the
index of the corresponding basis state but by its eigenvalue $\kappa$. Furthermore, consider a
window $1/N\ll\Delta\kappa\ll 1$ containing $m=\nu(\kappa)\Delta\kappa \gg 1$ levels. If $v_j$ do not
change appreciably within this window we may approximate its contribution to the left hand side of
Eq. (\ref{eq:swres}) by $4(v_j-v_i)/[\gamma(\kappa_j-\kappa_i)^2] \sum_{j\in\Delta\kappa}(H_{ij})^2$.
Since for $i\neq j$, $\mu[(H_{ij})^2]=1/(2N)$ and $\sigma[(H_{ij})^2]=1/(\sqrt{2}N)$ the central limit
theorem implies that $\sum_{j\in\Delta\kappa}(H_{ij})^2$ is normally distributed with mean $m/(2N)$ and
standard deviation $\sqrt{m}/(\sqrt{2}N)$. Consequently, we may neglect its fluctuations, replace it
by its mean and arrive at the following eigenvalue problem
\begin{equation}
\label{eq:eigsA2}
\frac{2}{\gamma N}P\int_{-\sqrt{2}}^{\sqrt{2}}\!d\kappa\,\nu(\kappa)\frac{v(\kappa)-v(\tau)}{(\kappa-\tau)^2}=\lambda v(\tau),
\end{equation}
where $P$ stands for the principle value of the integral. One can check by induction that the solutions of
this equation take the form
\begin{equation}
\label{eq:eigsAsol}
\lambda_n=-\frac{2}{\gamma}n,\;\;\;\;\;\; v_n(\kappa)=U_n\left(\frac{\kappa}{\sqrt{2}}\right),
\;\;\;\;\;\; n=0,1,2,\cdots
\end{equation}
where $U_n(x)$ are the Chebyshev polynomials of the second kind satisfying
\begin{equation}
\label{eq:Cheb}
U_0(x)=1, \;\;\; U_1(x)=2x, \;\;\; U_n(x)=2xU_{n-1}(x)-U_{n-2}(x).
\end{equation}

The eignvectors obey
\begin{eqnarray}
\label{eq:eigsprop1}
\int_{-\sqrt{2}}^{\sqrt{2}}\!d\kappa\,\nu(\kappa)v_m(\kappa)v_n(\kappa)&=&N\delta_{mn}\\
\label{eq:eigsprop2}
\int_{-\sqrt{2}}^{\sqrt{2}}\!d\kappa\,\nu(\kappa)v_n(\kappa)&=&N\delta_{0,n}.
\end{eqnarray}
Since they are used to expand the diagonal $\rho_{ii}$ of the density matrix, Eq. (\ref{eq:eigsprop2}) implies
that the zero mode $v_0$ carries a unit trace while the others are traceless. Hence, the zero mode must be
included in the expansion of a physical $\rho$ with coefficient $1/N$, while the expansion coefficients of the
remaining modes are free.

\bigskip

\begin{figure}[htbp]
\begin{center}
\includegraphics[width = 0.4\textwidth]{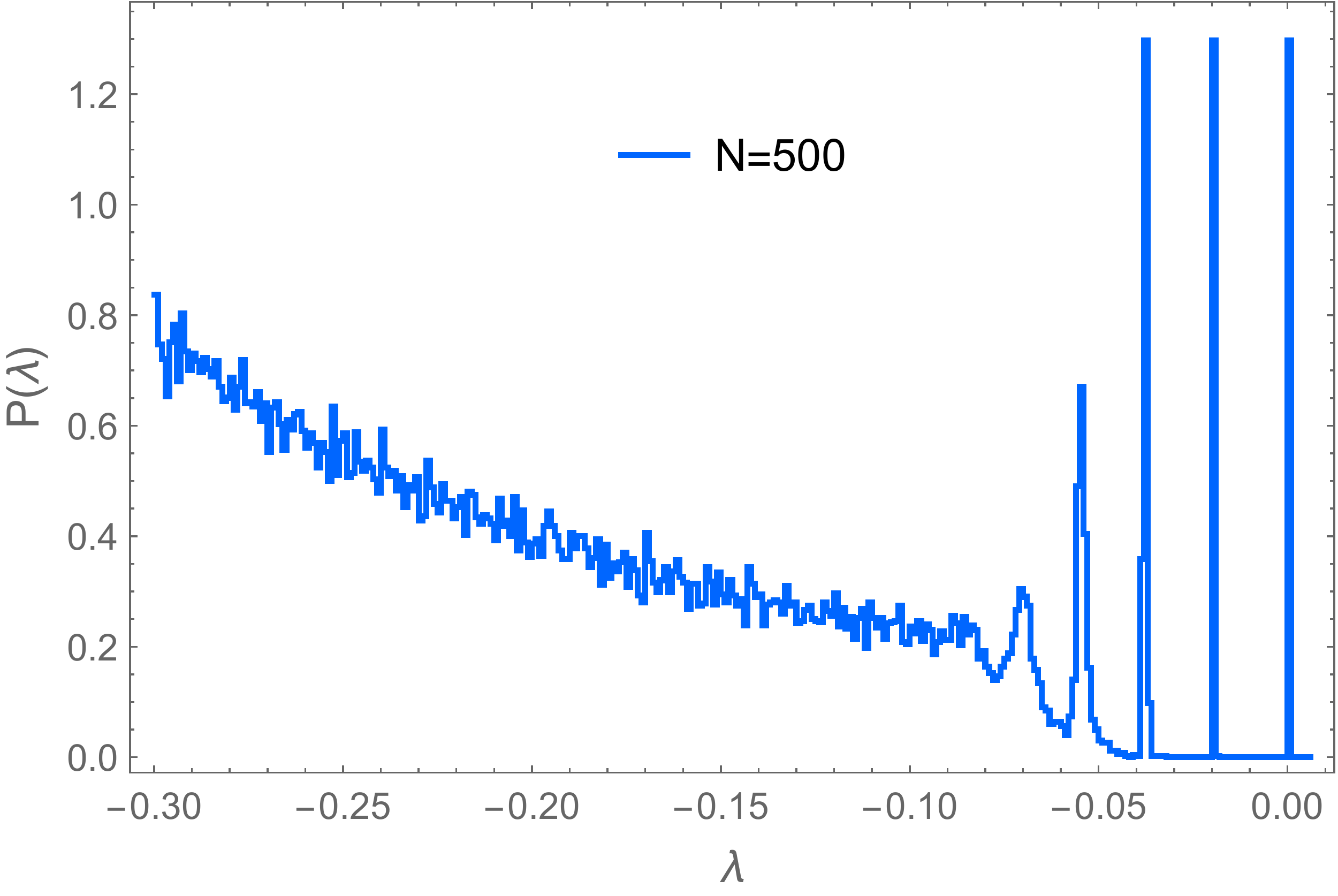}
\hspace{0.5cm}
\includegraphics[width = 0.4\textwidth]{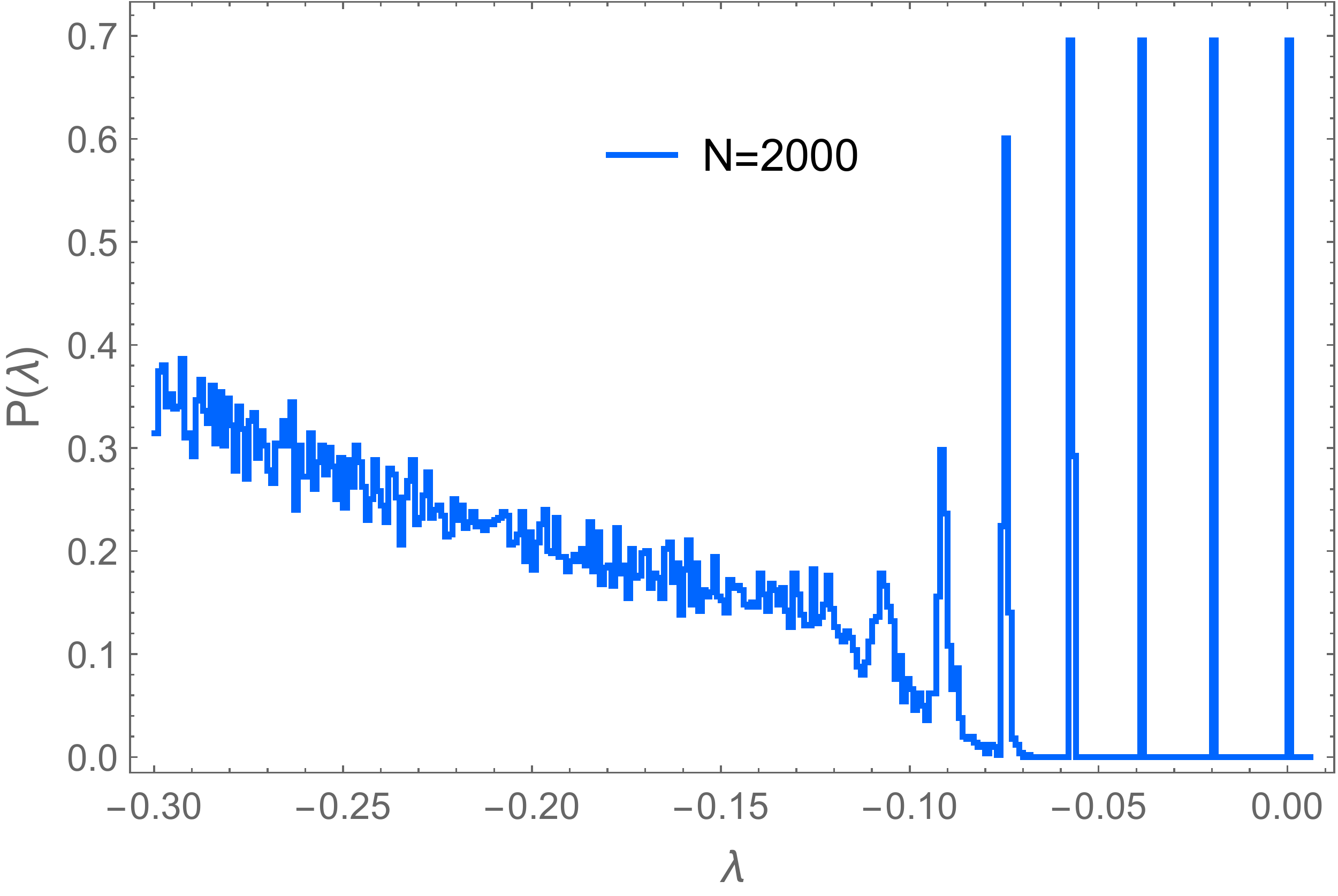}
\caption{Probability density of the eigenvalues of $A'$ for $\gamma = 100$, computed for two system sizes. 
At larger $N$ one sees more isolated eigenvalues.}
\label{sfig1}
\end{center}
\end{figure}

The above analysis relies on the assumption that the components of the eigenvectors do no change rapidly as 
function of $\kappa$. However, we note that  $U_n(\kappa)$ wiggles between $n$ zeros whose average separation 
across the support of the spectrum is $\sqrt{8}/n$. Moreover, changes are even faster near the edges of 
the spectrum where the separation between the zeros scales as $1/n^2$ and since $U_n(\pm 1)=n+1$. Thus, we 
expect growing deviations from the result, Eq. (\ref{eq:eigsAsol}), with increasing $n$. In order to check this we have 
numerically diagonalized $A'$. To make contact with the main text, we have done so for the effective model, which includes 
in $C_0$ also the first order correction to the eigenvalues of the $L$-coherences [see Eq. (4) of the main text]. 
We have checked that this does not affect the overall behavior. Representative results are shown in Fig.~\ref{sfig1}. 
We see that the spectrum of the reduced problem consists of a sequence of sharp peaks, which eventually merge into a continuum. 
Further, the scale (i.e., values of $\lambda$) at which a continuum forms is system-size dependent, with more isolated 
eigenvalues appearing as $N$ increases.


%

\newpage

\subsubsection{The spectrum of $F'$}

To estimate the eigenvalue distribution of $F'$ we treat its off-diagonal elements as a perturbation.
The diagonal elements are
\begin{equation}
\label{eq:PFpdiagel}
-\frac{\gamma}{2}(\kappa_i-\kappa_j)^2+i(H_{jj}-H_{ii})\equiv x_{ij}+iy_{ij}.
\;\;\;\;\;\;\;\;\;\;\;\;\;\;\; (i\neq j).
\end{equation}
Neglecting correlations between $\kappa_i$ and $\kappa_j$ we find that $x_{ij}$ is distributed according to
\begin{equation}
\label{eq:xijdist}
P(x_{ij})=\sqrt{\frac{2}{\gamma|x_{ij}|}}f\left(\sqrt{\frac{2|x_{ij}|}{\gamma}}\right)\Theta(-x_{ij})\Theta(4\gamma+x_{ij}),
\end{equation}
where $f(x)$ is given by Eq.(\ref{eq:fdef}), resulting in
\begin{equation}
\label{eq:xdistFp}
\mu(x)=-\frac{\gamma}{2},  \;\;\;\;\;\;\;\;
\sigma(x)=\sqrt{\frac{3}{8}}\gamma.
\end{equation}
Because of level repulsion Eq. (\ref{eq:xijdist}) needs to be modified for $|x_{ij}|<\gamma/N^2$, where the linear
$\kappa$-spacing distribution leads to $P(|x_{ij}|<\gamma/N^2)\sim N/\gamma$. The imaginary part, $y_{ij}$, is normally
distributed with
\begin{equation}
\label{eq:ymomentsFp}
\mu(y)=0,  \;\;\;\;\;\;\;\;
\sigma(y)=\sqrt{\frac{2}{N}}.
\end{equation}

Within second order perturbation theory the unperturbed eigenvalue $x_{ij}+iy_{ij}$ acquires a shift due to the
coupling of the corresponding eigenstate $\rho_{ij}$ to $2(N-2)$ other states $\rho_{kl}$ with either $k=i$ or $l=j$.
These couplings are of the form $iH_{jl}$, etc. As a result, the numerator of the corresponding perturbative term
\begin{equation}
\label{eq:pshiftFp}
\frac{-H_{jl}^2}{x_{ij}-x_{il}+i(y_{ij}-y_{il})}\equiv-\frac{\zeta_{ij\_ il}}{x_{ij\_ il}+iy_{ij\_ il}},
\end{equation}
is distributed according to
\begin{equation}
\label{eq:PzetaFp}
P_{\zeta}(\zeta)=\sqrt{\frac{N}{\pi}}\frac{1}{\sqrt{\zeta}}e^{-N\zeta}\,\Theta(\zeta).
\end{equation}
To estimate the shift in the real part of the eigenvalue due to these terms
\begin{equation}
\label{eq:realshiftFp}
\Delta x_{ij}=-\sum_{k\neq i}\frac{x_{ij\_ kj}}{x_{ij\_ kj}^2+y_{ij\_ kj}^2}\zeta_{ij\_ kj}-
\sum_{l\neq j}\frac{x_{ij\_ il}}{x_{ij\_ il}^2+y_{ij\_ il}^2}\zeta_{ij\_ il}
\equiv\sum_{k\neq i} \xi_{ij\_ kj}\zeta_{ij\_ kj}+\sum_{l\neq j} \xi_{ij\_ il}\zeta_{ij\_ il}
\equiv\sum_{k\neq i} \chi_{ij\_ kj}+\sum_{l\neq j} \chi_{ij\_ il},
\end{equation}
we approximate the function $f(x)$ in Eq. (\ref{eq:xijdist}) by a Gaussian with the same (zero) mean and (unit)
standard deviation as those of $f(x)$. As a result, and after neglecting correlations between $x_{ij}$ and $x_{il}$,
we obtain the following distribution of $x_{ij\_il}$
\begin{equation}
\label{eq:P(x)Fp}
P(x)=\Theta(x)\int_{-\infty}^0 dl \frac{e^{l/\gamma}}{\sqrt{-\pi\gamma l}}\frac{e^{(l-x)/\gamma}}{\sqrt{\pi\gamma(x-l)}}
+\Theta(-x)\int_{-\infty}^x dl \frac{e^{l/\gamma}}{\sqrt{-\pi\gamma l}}\frac{e^{(l-x)/\gamma}}{\sqrt{\pi\gamma(x-l)}}
=\frac{1}{\pi\gamma}K_0\left(\frac{|x|}{\gamma}\right).
\end{equation}
Taking into account the effect of level repulsion modifies the behavior at small $x$ leading to $P(|x|<1/N^2)\sim\ln(N)/\gamma$.
Noticing that $y_{ij\_il}=H_{jj}-H_{ll}$ we conclude that it is normally distributed with $\mu(y)=0$ and $\sigma(y)=\sqrt{2/N}$.
We then have for $\xi_{ij\_il}$
\begin{eqnarray}
\label{eq:P(xi)Fp}
\nonumber
P_{\xi}(\xi)&=&\frac{N^{1/2}}{2\pi^{3/2}\gamma}\int_{-\infty}^\infty dx dy\, K_0\left(\frac{|x|}{\gamma}\right)
e^{-Ny^2/4}\delta\left(\xi+\frac{x}{x^2+y^2}\right)\\
&=&\frac{N^{1/2}}{2\pi^{3/2}\gamma}\frac{1}{|\xi|^3}\int_{-\infty}^\infty dy \frac{1}{(1+y^2)^2}
K_0\left[\frac{1}{\gamma|\xi|(1+y^2)}\right]e^{-Ny^2/[4\xi^2(1+y^2)^2]},
\end{eqnarray}
which leads to the approximate behaviour
\begin{equation}
\label{eq:P(xi)approxFp}
P_{\xi}(\xi)\sim \left\{
\begin{array}{cc}
\frac{N^{1/2}}{\gamma}|\xi|^{-3} & \;\;\;\;\;\;|\xi|>N^{1/2} \\
\frac{1}{\gamma}|\xi|^{-2} & \;\;\;\;\;\;N^{1/2}>|\xi|>\gamma^{-1} \\
\frac{1}{\gamma N} & \;\;\;\;\;\;\gamma^{-1}>|\xi|
\end{array}
\right. .
\end{equation}
Using Eqs. (\ref{eq:PzetaFp}) and (\ref{eq:P(xi)approxFp}) we arrive at the distribution for $\xi_{ij\_il}$
\begin{equation}
\label{eq:P(chi)Fp}
P_{\chi}(\chi)=\int_0^\infty d\zeta\, \frac{1}{\zeta} P_{\zeta}(\zeta) P_{\xi}\left(\frac{\chi}{\zeta}\right)
\sim \left\{
\begin{array}{cc}
\frac{1}{\gamma N^{3/2}}|\chi|^{-3} & \;\;\;\;\;\;|\chi|>\frac{1}{N^{1/2}} \\
\frac{1}{\gamma N}|\chi|^{-2} &
\;\;\;\;\;\;\frac{1}{N^{1/2}}>|\chi|>\frac{1}{\gamma N} \\
\gamma^{1/2}N^{1/2}|\chi|^{-1/2}&
\;\;\;\;\;\;\frac{1}{\gamma N}>|\chi|
\end{array}
\right. .
\end{equation}
Once again, there is some numerical evidence that the decay in the range $1/N^{1/2}>|\chi|>1/(\gamma N)$
is slightly slower than $|\chi|^{-2}$ leading to $\mu(|\chi|)\sim 1/(\gamma N)$. For states near the upper edge
of the $x_{ij}$ distribution almost all of $x_{ij\_il}=x_{ij}-x_{il}$ are positive, and thus almost all
of the $2(N-2)$ perturbative corrections to the real part of their eigenvalue are negative, see Eq. (\ref{eq:realshiftFp}).
Consequently, the edge of the $x_{ij}$ distribution is shifted from zero by an amount of order $-1/\gamma$.
Numerically we find that for large $N$ the prefactor of the shift is larger than 2 and that the first few
eignevalues with the smallest real part (in terms of magnitude) come from the spectrum of $A'$.

\begin{figure}[!!!t]
\begin{center}
\begin{tabular}{cr}
\includegraphics[width = 0.37\textwidth]{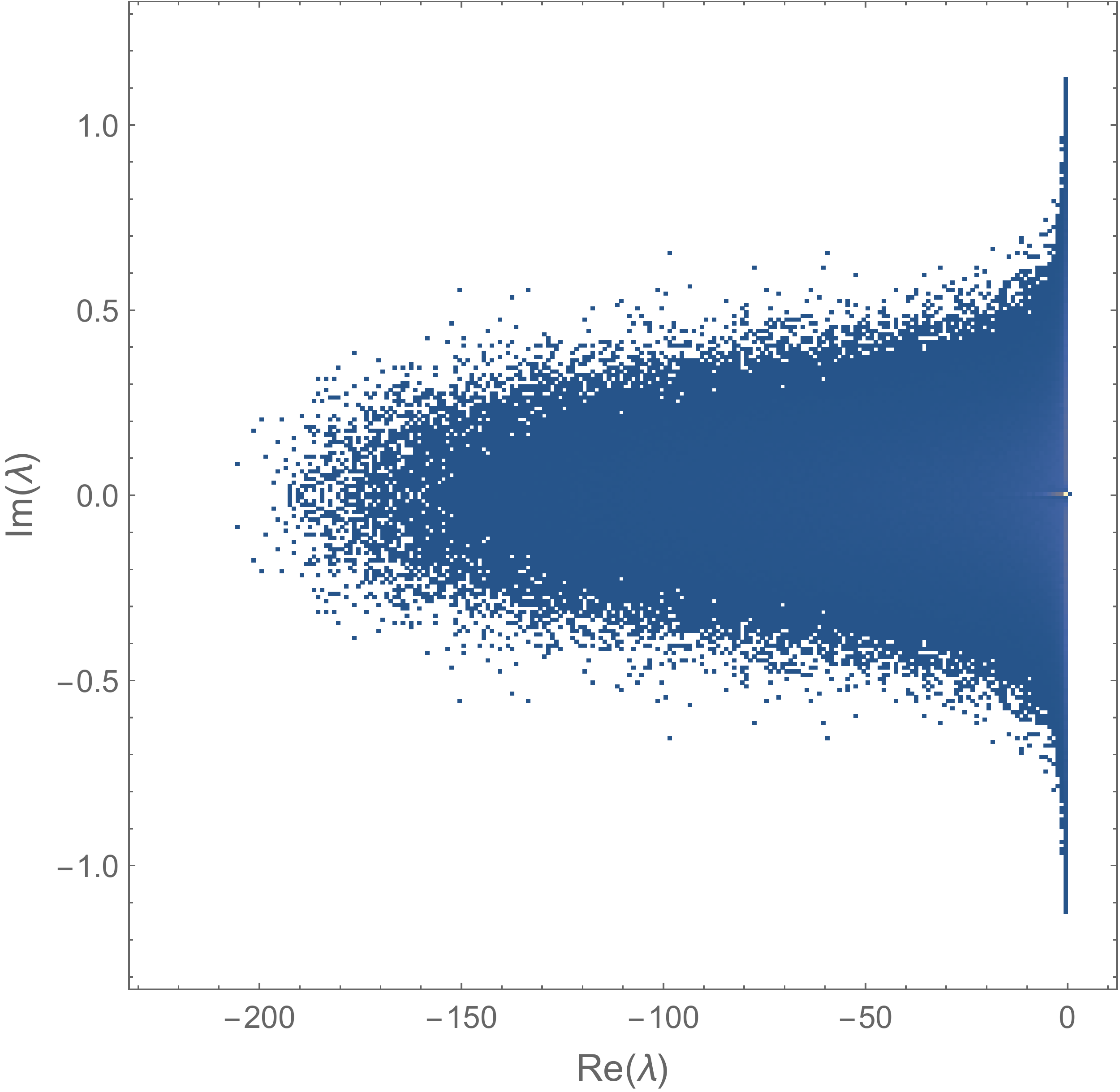} & \includegraphics[width = 0.53\textwidth]{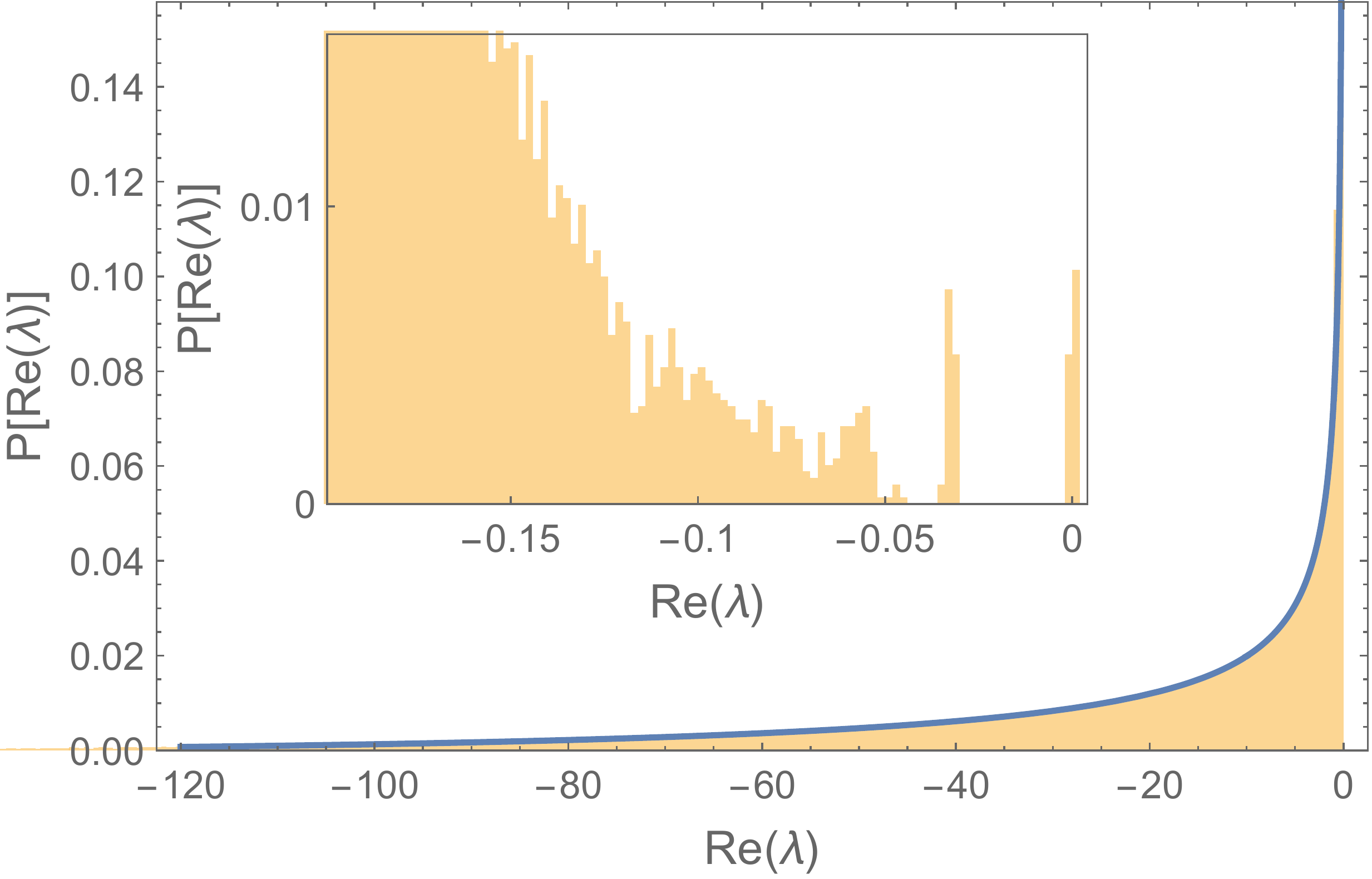} \\
\includegraphics[width = 0.37\textwidth]{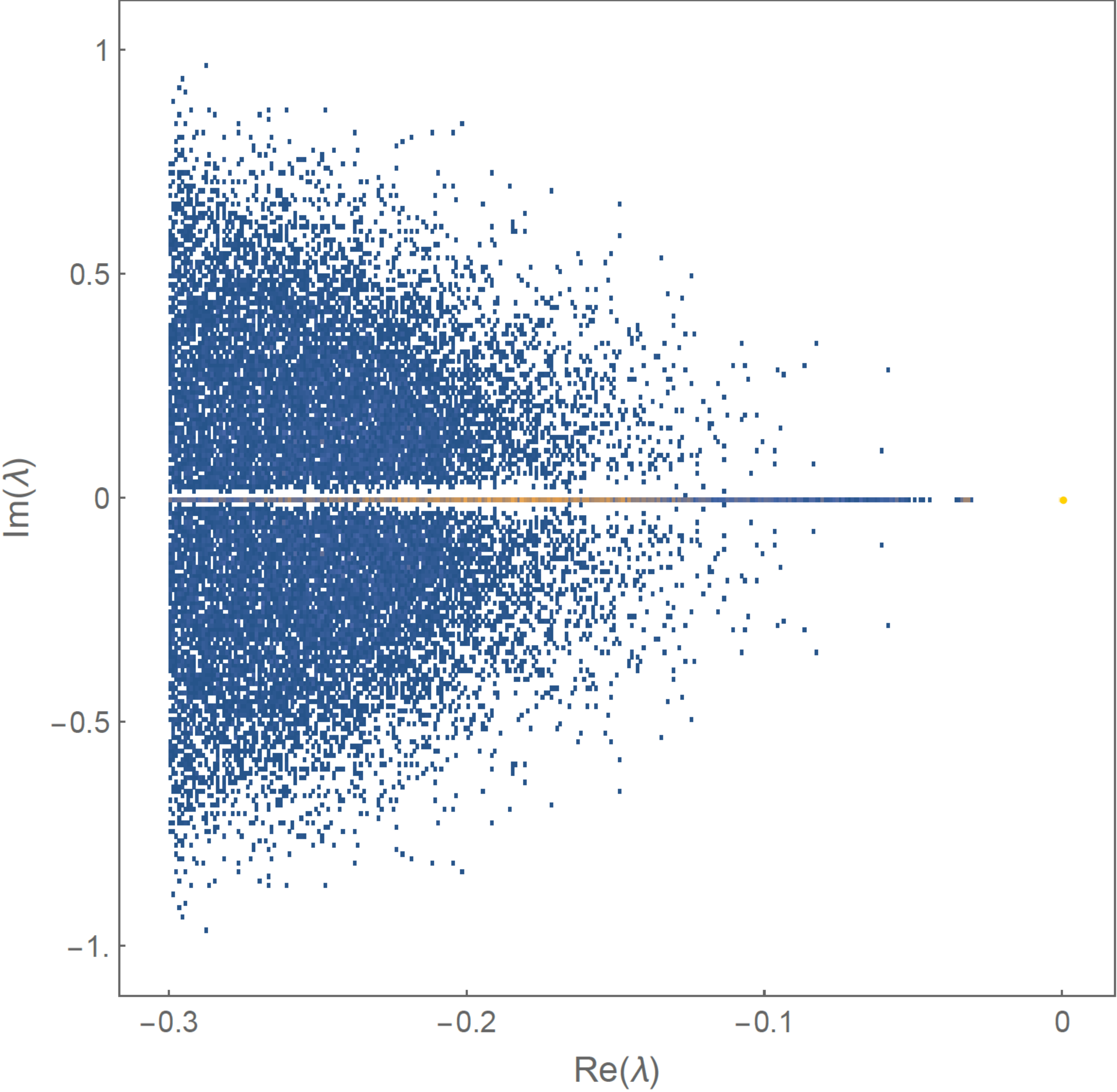} & \includegraphics[width = 0.515\textwidth]{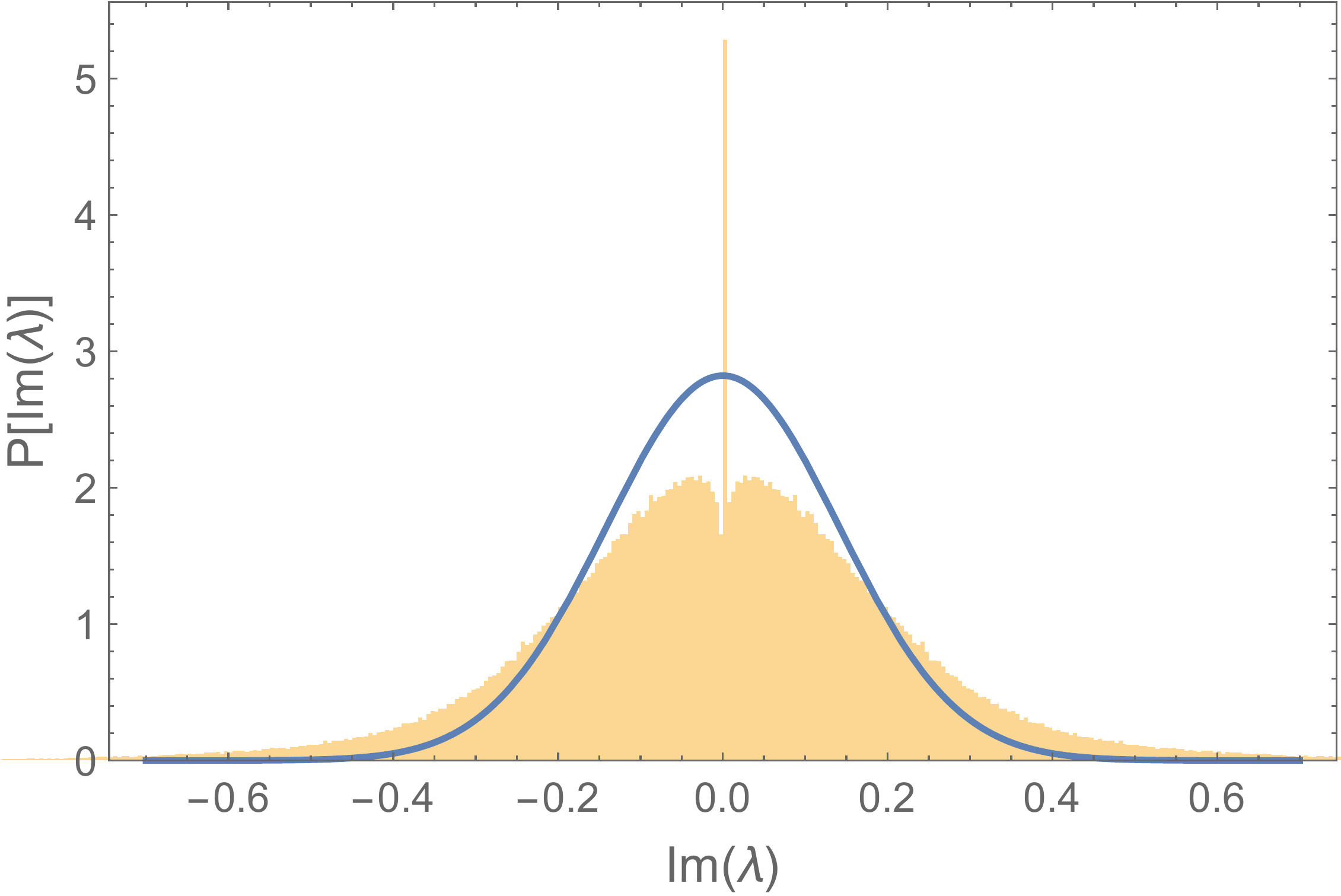}
\end{tabular}
\end{center}
\vspace{-0.5cm} \caption[]{\small Left: Eigenvalues distribution of $\L$, calculated from averaging over 60 realizations
for the case $\gamma=50$ and $N=100$. The lower panel depicts in more details the distribution near the origin.
Right: The upper panel contains the projection of the distribution on the real axis. The blue line corresponds to
Eq. (\ref{eq:xijdist}) and the inset depicts the edge of the distribution. Note the isolated state between the edge
of the eigenvalue cloud and the zero-mode at the origin. The lower panel shows the projection of the distribution on
the imaginary axis alongside a normal distribution with standard deviation $\sqrt{2/N}$. The latter corresponds
to the distribution of diagonal (unperturbed) elements of $\L$, and its deviation from the exact result is
due to eigenvalues with small real parts [it fits the data reasonably well for $\rm{Re}(\lambda)<-10$].
In the $N\rightarrow\infty$ limit we expect that the standard deviations of the real and imaginary parts
scale with $\gamma$ and $1/\gamma$, respectively.}
\label{fig-largegamma}
\end{figure}

\bigskip

For the shift in the imaginary part
\begin{equation}
\label{eq:imaglshiftFp}
\Delta y_{ij}=\sum_{k\neq i}\frac{y_{ij\_ kj}}{x_{ij\_ kj}^2+y_{ij\_ kj}^2}\zeta_{ij\_ kj}+
\sum_{l\neq j}\frac{y_{ij\_ il}}{x_{ij\_ il}^2+y_{ij\_ il}^2}\zeta_{ij\_ il}
\equiv\sum_{k\neq i} \eta_{ij\_ kj}\zeta_{ij\_ kj}+\sum_{l\neq j} \eta_{ij\_ il}\zeta_{ij\_ il}
\equiv\sum_{k\neq i} \nu_{ij\_ kj}+\sum_{l\neq j} \nu_{ij\_ il}.
\end{equation}
we need the distribution of $\eta_{ij\_il}$
\begin{eqnarray}
\label{eq:P(eta)Fp}
\nonumber
P_{\eta}(\eta)&=&\frac{N^{1/2}}{2\pi^{3/2}\gamma}\int_{-\infty}^\infty dx dy\, K_0\left(\frac{|x|}{\gamma}\right)
e^{-Ny^2/4}\delta\left(\eta-\frac{y}{x^2+y^2}\right)\\
&=&\frac{N^{1/2}}{2\pi^{3/2}\gamma}\frac{1}{|\eta|^3}\int_{-\infty}^\infty dx \frac{1}{(1+x^2)^2}
K_0\left[\frac{|x|}{\gamma|\xi|(1+x^2)}\right]e^{-N/[4\xi^2(1+x^2)^2]},
\end{eqnarray}
which can be approximated by
\begin{equation}
\label{eq:P(eta)approxFp}
P_{\eta}(\eta)\sim \left\{
\begin{array}{cc}
\frac{N^{1/2}}{\gamma}|\eta|^{-3} & \;\;\;\;\;\;|\eta|>N^{1/2} \\
\frac{1}{\gamma N^{1/4}}|\eta|^{-3/2} & \;\;\;\;\;\;N^{1/2}>|\eta|>\frac{1}{\gamma^2N^{1/2}} \\
\gamma^2 N^{1/2} & \;\;\;\;\;\;\frac{1}{\gamma^2N^{1/2}}>|\eta|
\end{array}
\right.
\end{equation}
and thus the distribution of the correction $\nu_{ij\_il}$ is
\begin{equation}
\label{eq:P(nu)Fp}
P_{\nu}(\nu)=\int_0^\infty d\zeta\, \frac{1}{\zeta} P_{\zeta}(\zeta) P_{\eta}\left(\frac{\nu}{\zeta}\right)
\sim \left\{
\begin{array}{cc}
\frac{1}{\gamma N^{3/2}}|\nu|^{-3} & \;\;\;\;\;\;|\nu|>\frac{1}{N^{1/2}} \\
\frac{1}{\gamma N^{3/4}}|\nu|^{-3/2} &
\;\;\;\;\;\;\frac{1}{N^{1/2}}>|\nu|>\frac{1}{\gamma^2 N^{3/2}} \\
\gamma^2 N^{3/2} &
\;\;\;\;\;\;\frac{1}{\gamma^2 N^{3/2}}>|\nu|
\end{array}
\right. .
\end{equation}
Using that Eq.(\ref{eq:P(nu)Fp}) results in $\mu(|\nu|)\sim 1/(\gamma N)$ we may estimate the average shift in the imaginary part
of the eigenvalues
\begin{equation}
\label{eq:ishiftapprxFp}
\mu(\Delta y_{ij})\sim \mu(|\nu|)\Bigg[\sum_{k\neq i}{\rm sign}(\nu_{ij\_ kj})+\sum_{l\neq j}{\rm sign}(\nu_{ij\_ il})\Bigg]
\sim \frac{1}{\gamma}{\rm erf}\left(\frac{y_{ij}}{2/\sqrt{N}}\right).
\end{equation}
A similar approximation to the one taken after Eq. (\ref{eq:rshiftapprx}) leads then to the conclusion that the shifted
imaginary parts $\tilde{y}_{ij}=y_{ij}+\Delta y_{ij}$ are normally distributed with $\mu(\tilde{y}_{ij})=0$ and
$\sigma(\tilde{y}_{ij})$ that varies from $\sqrt{2/N}$ for $N\ll\gamma^2$ to $O(1/\gamma)$ for $N\gg\gamma^2$.

\section{Small-$|\lambda|$ tails in other ensembles}

\begin{figure}[htbp]
\begin{center}
\includegraphics[width = 0.4\textwidth]{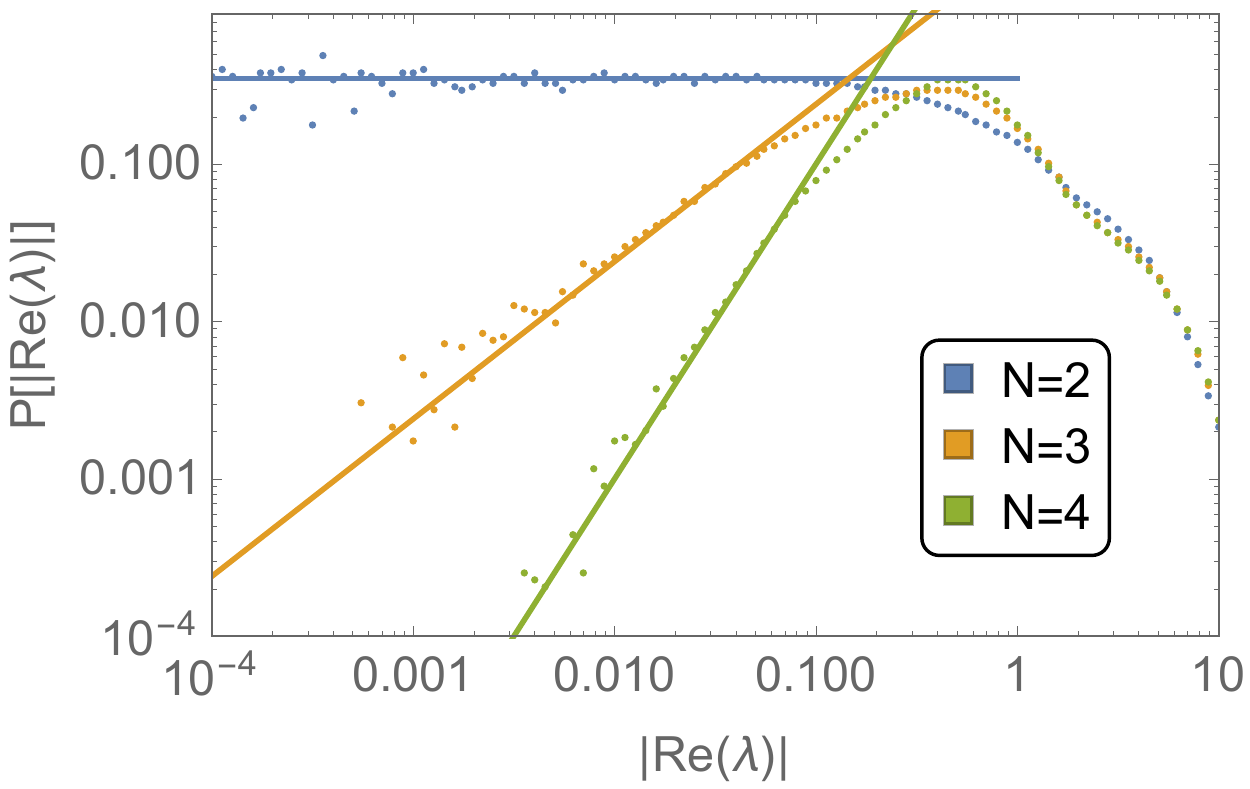}
\includegraphics[width = 0.4\textwidth]{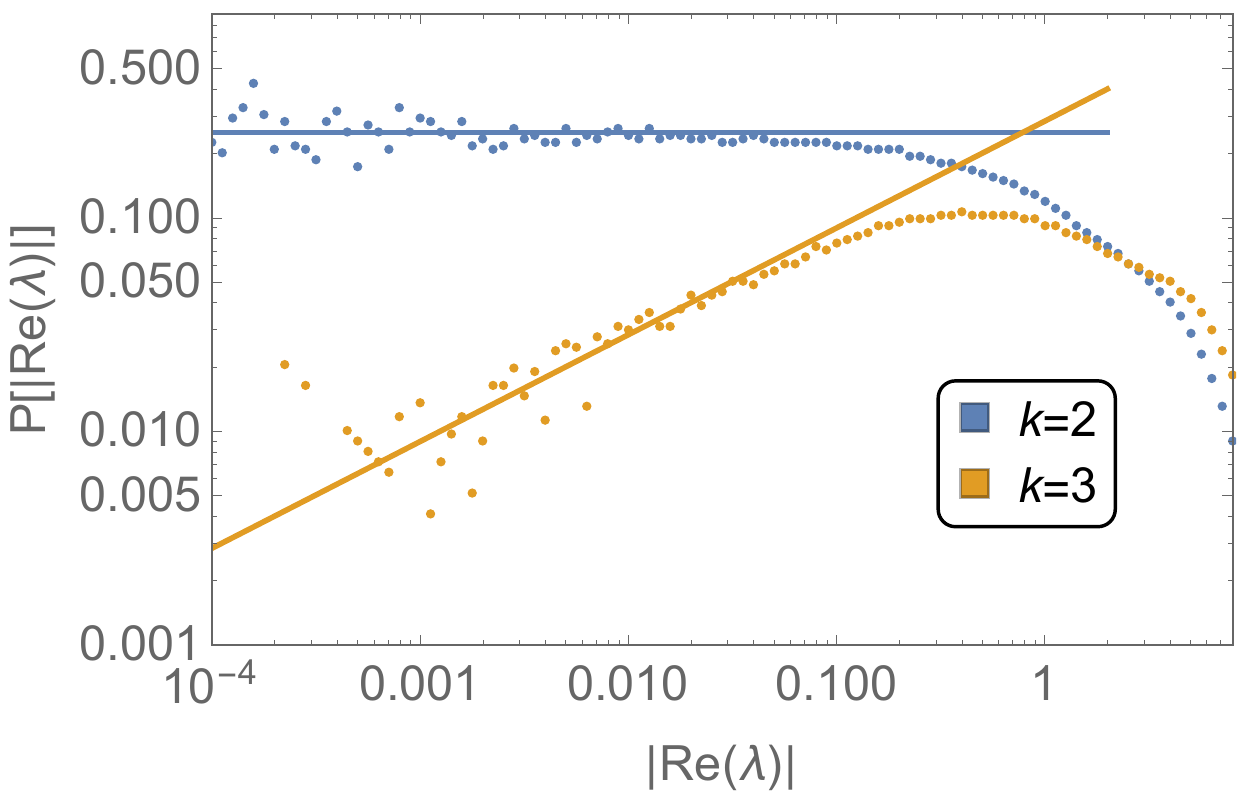}
\caption{Left: density of states at small $|\lambda|$ for Hamiltonians and dissipators chosen from the Gaussian unitary ensemble, for $\gamma = 2, k = 1$. Right: Data for the Gaussian orthogonal ensemble with $N=2$ and $k > 1$ distinct jump operators. Straight lines indicate the exponents according to Eq. (5) in the main text.}
\label{sfig2}
\end{center}
\end{figure}

In the main text we argued that the probability density of small gaps should obey a universal formula depending on the size $N$, the number of dissipators $k$, and the random-matrix ensemble $\beta$. In the main text we verified these predictions for the Gaussian orthogonal ensemble with a single jump operator. Here, we provide numerical support for this formula for the Gaussian unitary ensemble (i.e., matrices with complex entries) and for the case with $k > 1$ distinct jump operators. This numerical evidence is shown in Fig.~\ref{sfig2}: the naive predictions in the main text, based on counting independent random numbers, appear to work in all cases we have looked at. (We have also checked these results for the symplectic and Ginibre ensembles; these results will be presented elsewhere.)

